\documentclass[11pt,a4paper]{article}
\usepackage{jcappub}
\usepackage{epstopdf}
\usepackage{graphicx}
\usepackage{bm}
\usepackage{epstopdf}
\usepackage{mathtools}
\usepackage{natbib}
\usepackage{captcont}
\usepackage{booktabs}
\usepackage{threeparttable}
\usepackage{multirow}

\title{Testing clockwork axion with gravitational waves}

\author[a,b,1]{Cheng-Wei Chiang}
\author[a,2]{and Bo-Qiang Lu}

\affiliation[a]{Department of Physics, National Taiwan University, Taipei, Taiwan 10617, R.O.C.}
\affiliation[b]{Physics Division, National Center for Theoretical Sciences, Taipei, Taiwan 10617, R.O.C.}

\emailAdd{chengwei@phys.ntu.edu.tw}
\emailAdd{bqlu@phys.ntu.edu.tw}

\abstract{
We investigate the gravitational waves (GWs) produced from the Peccei-Quinn (PQ) phase transition associated with the clockwork axion.  The PQ phase transition can be first-order when the dimension-6 operator is included into the scalar potential.  The GWs from the PQ phase transition at scale in the range of $10^3-10^6$~GeV are detectable for the BBO and ALIA interferometers.  The LISA and Taiji interferometers can probe the GWs from the PQ scale $f\lesssim 10^4$~GeV, while the GW signals from the scale $f\gtrsim 10^5$~GeV can be detected by the ground-based GW observatories ET and CE.  We find that the parameter space $\kappa_m\sim 0.06-0.001$, $\kappa_l\sim 0.04-0.001$, and $\varepsilon\sim 0.1-0.01$ at the scale $f=10^5$~GeV and most of the parameter regions at the scale $f=10^6$~GeV 
have been excluded by the LIGO O2 run. The LIGO O3 and design phases can further probe the remaining parameter space.  We show that the GWs from the annihilation of domain walls with a PQ scale $f\simeq 2\times 10^5$~GeV can induce the stochastic signals indicated by the 12.5-year observation of NANOGrav.  The LIGO O3 run has the opportunity of detecting the GW signals from the first-order PQ phase transition around this scale.
}


\arxivnumber{}

\begin{document}
\maketitle

\section{Introduction}

First-order cosmological phase transitions are of particular interest because these violent phenomena in the early Universe can be the production source of gravitational waves (GWs), which can be probed by current and future GW experiments~\cite{Schwaller2015PRL}.
The phase transitions are relevant to the spontaneous breakdown of symmetries in particle physics. As the Universe temperature drops to the symmetry breaking scale, the vacuum of the Universe transits from a symmetric phase to a broken one.  Within the Standard Model (SM) of particle physics, there are two phase transitions: the electroweak weak phase transition (EWPT) at $\sim 100$~GeV and the QCD phase transition at $\sim 0.1$~GeV.  However, both of them are found to be crossovers rather than first-order phase transitions~\cite{Onofrio2014PRL,Aoki2016Nature}. Extensions of the SM to render the EWPT first-order have been widely studied in the literature~
\cite{Espinosa2008PRD,Barger2008PRD,Barger2009PRD,Espinosa2012NPB,Li2014JHEP,Chiang2014PLB,Profumo2015PRD,Kotwal2016PRD,Beniwal2017JHEP,Cline2017PRD,Alves2019JHEP,Ghosh2020,
Gould2019PRD,Kozaczuk2020PRD,Jiang2016PRD,Chiang2018PRD,Niemi2019PRD,Chao2019JHEP,Basler2017JHEP,Dorsch2017JHEP,Bernon2018JHEP,Andersen2018PRL,
Huang2016PRD,Huang2017PRD,Chala2018JHEP,Grzadkowski2018JHEP,Carena2020JHEP,Musolf2020JHEP,Grojean2005PRD,Cai2017JCAP,Chiang2020JHEP,Chiang2020JCAP}.
Future space-based interferometers can be used to test these models since the GW signals from the EWPT peak around the mHz range.

In addition to the SM symmetries, the solution to the strong CP problem via the Peccei-Quinn (PQ) mechanism~\cite{Peccei1977PRL,Peccei1977PRD} demands the existence of a global $U(1)_{\rm PQ}$ symmetry. 
The QCD axion~\cite{Weinberg1978PRL,Wilczek1978PRL,Kim1979PRL,Shifman1980NPB,Dine1981PLB,Zhitnitsky1980SJNP}, 
the pseudo-Goldstone boson associated with the PQ symmetry breaking, can serve as an attractive dark matter candidate.  In the conventional QCD axion scenarios, the axion decay constant $f_a$ is at the same order as the PQ symmetry breaking scale $f$, {\it i.e.,} $f_a\sim f$.  The axion decay constant has been restricted to the range $10^{9}\lesssim f_a\lesssim 10^{12}$~GeV (see, for example, refs.~\cite{Marsh2016PR,Luzio2020PR} for recent reviews on the QCD axion).  The lower bound comes from the the SN~1987A neutrino burst duration observations~\cite{Mayle1988PLB,Raffelt1988PRL,Turner1988PRL}, while the upper bound is to ensure that the Universe is not over-closed by the axion dark matter~\cite{Preskill1983PLB,Abbott1983PLB,Dine1983PLB}.  As a consequence, the classical QCD axion is nearly invisible since the QCD axion-gluon coupling is inversely proportional to the axion decay constant.  However, it is not necessary to associate the axion decay constant close to the PQ scale.  The canonical association of the axion decay constant to the PQ scale can be circumvented with the help of 
clockwork mechanism~\cite{Kaplan2016PRD}. In the clockwork axion model~\cite{Kaplan2016PRD,Choi2016JHEP,Higaki2016JHEPa,Higaki2016JHEPb,Giudice2017JHEP,Coy2017JHEP,Long2018JHEP,Agrawal2018JHEP}, $N+1$ complex scalar fields with global $U(1)$ symmetries are 
introduced and the axion decay constant can be exponentially enlarged with respect to the symmetry breaking scale~\cite{Kaplan2016PRD}. 
The clockwork mechanism allows a PQ symmetry breaking scale $f\lesssim 10^9$~GeV while keeping the axion decay constant $f_a$ consistent with the cosmological/astrophysical observations.

The abundant particle and cosmology phenomena for $f\ll f_a$ in the clockwork axion model have been investigated in the literature~\cite{Higaki2016JHEPa,Higaki2016JHEPb,Long2018JHEP,Agrawal2018JHEP}.  In this work, we are concerned with testing the clockwork axion model with the GW observations.  This is possible especially when the phase transition of the PQ symmetry breakdown is first-order.  However, the PQ phase transition in the conventional QCD axion models is only second-order.  Attempts to make a first-order PQ phase transition have been made in recent works~\cite{Croon2019JHEP,Dev2019JCAP,Harling2020JHEP,Rose2020JHEP,Ghoshal2020}.  One of the simplest scenarios is to introduce one or more Higgs(-like) doublets and the PQ complex scalar is coupled to the Higgs fields via the renormalization operator $\lambda_m|\Phi|^2|H|^2$~\cite{Dev2019JCAP,Harling2020JHEP}.  It is found that to obtain a first-order PQ phase transition, one needs a large Higgs portal coupling $\lambda_m\gtrsim 1$ and a small PQ scalar self-coupling $\lambda\sim 10^{-3}$~\cite{Dev2019JCAP}, which may be confronted with the constraints from the Higgs properties~\cite{Rose2020JHEP}.  The realization of a first-order PQ phase transition may also be achieved in the radiative PQ symmetry breaking scenario~\cite{Rose2020JHEP,Ghoshal2020} or in the composite axion models~\cite{Rose2020JHEP}.  In this work, to make the PQ phase transition first-order, we will introduce in the scalar potential a dimension-6 operator that can be generated by decoupling a massive degree of freedom.  This scenario has been investigated in the context of first-order EWPT in the literature~\cite{Grojean2005PRD,Huang2016PRD,Cai2017JCAP,Chala2018JHEP,Ellis2019JCAP,Musolf2020JHEP}.  We will show that the future space-based interferometers, such as, LISA~\cite{LISA2017,LISA2019CQG}, Taiji~\cite{Hu2017NSR,Ruan2020NA}, ALIA~\cite{ALIA2014JPCS}, DECIGO~\cite{DECIGO2017}, and BBO~\cite{BBO2006CQG}, and the ground-based GW observatories including Einstein Telescope (ET)~\cite{ET2010CQG}, Cosmic Explorer (CE)~\cite{CE2017CQG}, and Advanced LIGO (aLIGO)~\cite{LIGO2019,LIGOSGW2019PRD,LIGOSGW2018PRL} can explore the PQ symmetry breaking scale of the clockwork axion model in a broad range of $10^{3}-10^{6}$~GeV.  For the clockwork axion models with a PQ scale $f\gtrsim 10^{6}$~GeV, the domain walls produced from the phase transition would dominate the energy density of the Universe and have therefore been excluded.  We find that GWs produced from the annihilation of domain walls with a PQ scale $f\simeq 2\times 10^5$~GeV can account for the signal in the stochastic GW background from the analysis of 12.5-year data collected by the North American Nanohertz Observatory for Gravitational Waves (NANOGrav)~\cite{NANOGrav2020}.  For the phase transition at the scale $f\simeq 2\times 10^5$~GeV, we also expect to find a footprint of the clockwork axion on the stochastic GW background in the LIGO O3 run~\cite{LIGOSGW2019PRD,LIGOSGW2018PRL}.

This work is presented as follows.  In Sec.~\ref{sec:CWaxion}, we briefly review the clockwork axion model.  In Sec.~\ref{sec:pt}, we study in detail the phase transition in our model and perform a scan of model parameter space. The nucleation temperature of the true vacuum and the GW parameters are calculated in Sec.~\ref{sec:bn}.  The spectrum of GWs coming from the first-order phase transition and the annihilation of domain walls along with their detections in GW experiments are analyzed in Sec.~\ref{sec:GWPT} and Sec.~\ref{sec:GWDW}, respectively.  Finally in Sec.~\ref{sec:summary}, we summarize our findings.

\section{The clockwork axion model}
\label{sec:CWaxion}

In this section we briefly review the clockwork axion model, which has been widely studied in refs.~\cite{Kaplan2016PRD,Choi2016JHEP,Higaki2016JHEPa,
Higaki2016JHEPb,Giudice2017JHEP,Coy2017JHEP,Long2018JHEP,Agrawal2018JHEP}.
The clockwork model contains a number of $N+1$ complex scalars, denoted as $\Phi_i(x)$ with $i=0,1,...,N$.
The potential of these scalars are determined by
\begin{equation}
    \label{eq:poten1}
    V(\Phi)=\sum_{j=0}^{N}\left(-m^{2}\left|\Phi_{j}\right|^{2}+\frac{\lambda}{4}\left|\Phi_{j}\right|^{4}\right)-
    \varepsilon \sum_{j=0}^{N-1}\left( \Phi_{j}^{\dagger} \Phi_{j+1}^{3}+\rm h.c.\right)
    ~,
\end{equation}
where the parameters $m^2$, $\lambda$, and $\varepsilon$ have been assumed to be real and universal.  The first term respects a global $U(1)^{N+1}$ symmetry, which is explicitly broken by the $\varepsilon$-dependent term down to a 
global $U(1)$ symmetry
\begin{equation}
    \mathrm{U}(1): \Phi_{i} \rightarrow \exp \left[i q^{N-i} \theta\right] \Phi_{i},
\end{equation} 
with $0\leq \theta <2\pi$ and $q\equiv 3$. The global $U(1)$ symmetry is identified as the PQ symmetry in the clockwork work axion model.

Even without the $\varepsilon$-dependent term, the global $U(1)$ symmetry of the potential could be spontaneously broken when the radial 
components of the $N+1$ complex scalars acquire a nonzero  vacuum expectation value (VEV) $\left \langle \Phi_i \right \rangle=f/\sqrt{2}$, 
where $f$ is the $U(1)$ symmetry breaking scale and is assumed to be the same for all $\Phi_i$. Since now the $U(1)^{N+1}$ symmetry is 
explicitly broken to $U(1)$ by the $\varepsilon$-dependent term, the spontaneous symmetry breaking of the potential \eqref{eq:poten1} leads 
to $N$ massive pseudo-Goldstone bosons and one massless Goldstone boson.  
After the spontaneous symmetry breaking, we parametrize the scalar field as $\Phi_i=fe^{i\pi_i/f}/\sqrt{2}$ and obtain the potential for 
the $N+1$ Goldstone bosons
\begin{equation}
    \label{eq:Vpi}
    V(\pi)=-\frac{1}{2} \varepsilon f^{4} \sum_{i=0}^{N-1} \cos \frac{\pi_{i}-q \pi_{i+1}}{f} \simeq 
    \frac{\varepsilon f^{2}}{4} \sum_{i=0}^{N-1}\left(\pi_{i}-q \pi_{i+1}\right)^{2}
    =\frac{1}{2}\sum_{i,j=0}^{N}\pi_j \left( M_{\pi}^2 \right)_{ji} \pi_i
    ~,
\end{equation}
where the constant term is omitted and the mass matrix $M_{\pi ij}^2$ is given by
\begin{equation}
    M_{\pi}^{2}=m_{G}^{2}\left(\begin{array}{cccccc}
    1 & -q & 0 & \cdots & & 0 \\
    -q & 1+q^{2} & -q & \cdots & & 0 \\
    0 & -q & 1+q^{2} & \cdots & & 0 \\
    \vdots & \vdots & \vdots & \ddots & & \vdots \\
    & & & & 1+q^{2} & -q \\
    0 & 0 & 0 & \cdots & -q & q^{2}
    \end{array}\right),
\end{equation}
where $m_G^2=\varepsilon f^2/2$. One then rotates the $\pi_i$ fields to the mass eigenstate $a_i \equiv (a, A_1, \dots, A_N)$ by a real $(N+1)\times (N+1)$ orthogonal matrix $O$ so that 
the the mass matrix is diagonalized as
$O^{T} M_{\pi}^{2} O=\operatorname{diag}\left(m_{a}^{2},m_{A_1}^2, \ldots, m_{A_{N}}^{2}\right)$, 
where the eigenvalues of $N+1$ Goldstone bosons $a_i$ are given by
\begin{equation}
    \label{eq:massG}
    m_{a}^2=0~{\rm and}~m_{A_k}=\eta_km_{G}^2
    ~\mbox{with}~ \eta_{k}\equiv q^{2}+1-2q\cos\frac{k\pi}{N+1}
    ~\left( k=1,2,...,N \right)
    ~.
\end{equation} 
The massless Goldstone boson $a$ is identified as the axion and the $N$ massive pseudo-Goldstone states $A_k$ are the so-called gear 
fields since they play the role of `gears' in the clockwork mechanism.
The matrix elements of $O$ are given by
\begin{equation}
    \label{eq:rote0}
    O_{i 0}=\frac{\mathcal{N}_{0}}{q^{i}}, \quad O_{i k}=\mathcal{N}_{k}\left[q \sin \frac{i k \pi}{N+1}-\sin \frac{(i+1) k \pi}{N+1}\right],
\end{equation}
with $i=0,1,...,N$, $k=1,2,...,N$ and
\begin{equation}
    \label{eq:rote1}
    \mathcal{N}_{0} \equiv \sqrt{\frac{q^{2}-1}{q^{2}-q^{-2 N}}}
    ~, \quad 
    \mathcal{N}_{k} \equiv \sqrt{\frac{2}{(N+1) \eta_{k}}}
    ~.
\end{equation}

The $(N+1)$ $a_i$ fields are related to the $\pi_i$ fields by the rotation
\begin{equation}
    \label{eq:piN}
    \pi_i=\sum_{j=0}^{N}O_{ij}a_{j}
    \equiv O_{i0}a+\sum_{j=1}^{N}O_{ij}A_j
    ~.
\end{equation}
The potential of the (pseudo-)Goldstone bosons in the physical basis is then given by the sum of the contributions from all sites
\begin{equation}
    \label{eq:gearV}
    V(\pi)=\sum_{j=0}^{N} V_j(A_j)=\frac{1}{2}m_{G}^2\sum_{j=1}^{N}\eta_jA_j^2=\frac{1}{4}\varepsilon f^2\sum_{j=1}^{N}\eta_jA_j^2
    ~.
\end{equation}
Here we have used the fact that $m_a^2=0$.

The clockwork mechanism is illustrated as follows. 
Consider the effective Lagrangian in which the $N$-th site $\pi_N$ is coupled to the QCD topological term
\begin{equation}
    \label{eq:qcdtopol}
    \mathcal{L}\supset\frac{\alpha_{s}}{8 \pi}\frac{\pi_{N}}{f} G_{\mu \nu}^{a} \tilde{G}^{\mu \nu, a}.
\end{equation}
where $G_{\mu \nu}^{a}$ is the gluon field strength tensor, as seen in the Kim-Shifman-Vainshtein-Zakharov (KSVZ)~\cite{Kim1979PRL,Shifman1980NPB} type and 
the Dine-Fischler-Srednicki-Zhitnitsky (DFSZ)~\cite{Dine1981PLB,Zhitnitsky1980SJNP} type of axion models. 
Using eq.~\eqref{eq:piN}, the axion coupling to the topological term is then given by
\begin{equation}
    \mathcal{L}\supset\frac{\alpha_{s}}{8 \pi}\frac{a}{f_{a}} G_{\mu \nu}^{a} \tilde{G}^{\mu \nu, a}
    ~,
\end{equation}
where we have defined
\begin{equation}
    \label{eq:decayCont}
    f_a \equiv \frac{f}{O_{N0}}=\frac{q^Nf}{\mathcal{N}_{0}}\simeq q^Nf
    ~.
\end{equation} 
If the QCD topological term~\eqref{eq:qcdtopol} occurs at the `first' site $i=N$, we observe from eq.~\eqref{eq:rote0} that the coupling of the massless axion at the `last' site $i=0$ is suppressed by a factor of $q^N$.  In other words, the axion decay constant $f_a$ is amplified by a factor of $q^N$ compared to the symmetry breaking scale $f$, as given in eq.~\eqref{eq:decayCont}.
With the clockwork mechanism, a low PQ symmetry breaking scale $f$ and a nearly invisible axion can be simultaneously achieved in an axion model.

\section{Phase transition}
\label{sec:pt}

In this section, we discuss the phase transition associated with the PQ symmetry breakdown.  Various realizations of a first-order phase transition of the global $U(1)$ symmetry (PQ symmetry) breaking have been discussed in 
refs.~\cite{Croon2019JHEP,Dev2019JCAP,Harling2020JHEP,Rose2020JHEP,Ghoshal2020}, which include adding one~\cite{Dev2019JCAP} or 
two~\cite{Harling2020JHEP} Higgs(-like) doublets to the scalar potential, radiative PQ symmetry breaking~\cite{Rose2020JHEP,Ghoshal2020},
and the composite axion models~\cite{Rose2020JHEP}. In this work, we make the first-order phase transition for the global $U(1)$ symmetry 
breaking possible by adding a dimension-6 operator to the scalar potential
\begin{equation}
    \label{eq:poten2}
    V_{\Lambda}(\Phi)=V(\Phi)+\sum_{j=0}^{N}\frac{1}{\Lambda^2}\left|\Phi_{j}\right|^{6},
\end{equation}
where $\Lambda \geq f$ is the cut-off scale of the theory.

\subsection{The effective potential}

Let's first pay attention to the vacuum phase transition at the $j$-th site only.  The complex scalar $\Phi_j$ can be expanded around the classical backgrounds as
\begin{equation}
    \Phi_j=\phi_j e^{-i\pi_j/f}/\sqrt{2}
    ~.
\end{equation}
At finite temperature, the effective one-loop scalar potential is given by
\begin{equation}
    \label{eq:Veff}
    V_{\rm eff}(\phi,T)=V_0(\phi)+V_{\rm CW}(\phi)+V_T(\phi,T)+V_{\rm ring}(\phi,T)
    ~.
\end{equation}
Following ref.~\cite{Kaplan2016PRD}, here we have assumed the radial field $\phi_j \equiv \phi$, which develops a VEV $\langle \phi \rangle=f$ after the spontaneously $U(1)$ 
symmetry breaking. The tree-level scalar potential at zero temperature is given by
\begin{equation}
    V_0(\phi)= -\frac{1}{2}m^2\phi^2+\frac{1}{4}\lambda \phi^4 +\frac{1}{\Lambda^2}\phi^6.
\end{equation}
By requiring the renormalization conditions
\begin{equation}
    V_{\mathrm{CW}}(\phi=f)=V_{\mathrm{CW}}^{\prime}(\phi=f)=V_{\mathrm{CW}}^{\prime \prime}(\phi=f)=0
    ~,
\end{equation}
the Coleman-Weinberg potential can be written as~\cite{Coleman1973PRD}
\begin{equation}
    V_{\mathrm{CW}}(\phi)=\sum_{i} \frac{n_{i}}{64 \pi^{2}}\left[m_{i}^{4}(\phi)\left(\log \frac{m_{i}^{2}(\phi)}{m_{i}^{2}(f)}-
    \frac{3}{2}\right)+2 m_{i}^{2}(\phi) m_{i}^{2}(f)-\frac{m_{i}^{4}(\phi)}{2}\right]
    ~,
\end{equation}
where the subscript $i=\{ \phi, A \}$, where $A = A_j$ is the $j$-th site gear field whose potential is given by eq.~\eqref{eq:gearV},
and the number of degrees of freedom $n_i=\{ 1,1 \}$. The gear contributes to the effective potential at loop level.
The gear's mass depends on its site (see eq.~\eqref{eq:massG}), with the value of $\eta_j$ falling in the range of $\sim 4-16$.
Conservatively, we assume $\eta_j\equiv \eta\simeq 4$ for the gears to simplify the estimation.  For small values of $\phi$, the field-dependent mass of $\phi$ is negative, leading to a complex effective potential.  The imaginary part of the effective potential is related to the decay rate of the scalar~\cite{EJWeinberg1987PRD}.  This part can be abandoned in the calculation of phase transition since it is found to be tiny compared to the real part around the transition temperature~\cite{Delaunay2008JHEP}.

The finite-temperature contributions to the effective potential at one-loop level are given by~\cite{Dolan1974PRD}
\begin{equation}
    V_{\mathrm{T}}(\phi, T)=\frac{T^{4}}{2 \pi^{2}} \sum_{i} n_{i} J_{B, F}\left(m_{i}^{2}(\phi)/T^{2}\right)
    ~,
\end{equation}
where the thermal functions are defined as
\begin{equation}
    J_{B, F}\left(z^{2}\right)=\int_{0}^{\infty} d x x^{2} \ln \left(1 \mp e^{-\sqrt{x^{2}+z^{2}}}\right),
\end{equation}
with the minus sign for bosons ($B$) and the plus sign for fermions ($F$).  The ring diagram part of one-loop finite-temperature potential from bosons is given by 
\begin{equation}
    V_{\text {ring }}(\phi, T)=\sum_{i} \frac{n_{i} T}{12 \pi}\left[m_{i}^{3}(\phi)-\left(m_{i}^{2}(\phi)+
    \Pi_{i}(T)\right)^{\frac{3}{2}}\right]
    ~,
\end{equation}
where 
\begin{equation}
    \Pi_{\phi}(T)=\left( \frac{m_{\phi}^2}{4f^2}+\frac{2}{3}\epsilon-\frac{3f^2}{4\Lambda^2} \right )T^2~{\rm and}~
    \Pi_{A}(T)=\frac{2}{3}\epsilon T^2
    ~.
\end{equation}

\subsection{The high temperature expansion}
\label{sec:highT}

To get analytic insights on the phase transition associated with potential~\eqref{eq:poten2}, we study in this section the high temperature expansion of the potential given by 
\begin{equation}
    \label{eq:vht1}
    V_{\rm HT}(\phi,T)= -\frac{1}{2}m^2\phi^2+\frac{1}{4}\lambda \phi^4 +\frac{1}{\Lambda^2}\phi^6 +\frac{1}{2}CT^2\phi^2 
    ~,
\end{equation}
where 
\begin{equation}
    C=\frac{m_{\phi}^2}{4f^2}+\frac{2}{3}\epsilon-\frac{3f^2}{4\Lambda^2}
    ~.
\end{equation}
Using the renormalization conditions
\begin{equation}
    V_{\rm HT}^{\prime}(f,0)=0 \quad {\rm and} \quad V_{\rm HT}^{\prime \prime}(f,0)=m_{\phi}^{2}
    ~,
\end{equation}
where the prime denotes the derivation with respect to $\phi$ and $m_{\phi}$ is the mass of $\phi$.  We thus have
\begin{equation}
    m^{2}=\frac{m_{\phi}^{2}}{2}-\frac{3 f^{4}}{4 \Lambda^{2}}, \quad \lambda=\frac{m_{\phi}^{2}}{2 f^{2}}
    -\frac{3 f^{2}}{2 \Lambda^{2}}
    ~.
\end{equation}
The signs of $m^2$ and $\lambda$ depend on two parameters
\begin{equation}
    \kappa_{m}=\frac{m_{\phi}}{f} ~~{\rm and}~~ \kappa_{l}=\frac{f}{\Lambda}
    ~.
\end{equation}
From the potential~\eqref{eq:vht1} we see that at zero temperature, the tree-level barrier that separates the false and true vacua could 
arise by requiring that both $m^2$ and $\lambda$ be negative. We thus have
\begin{equation}
    \label{eq:bound1}
    \kappa_m<\sqrt{\frac{3}{2}}\kappa_l
    ~.
\end{equation}
This bound can be alleviated when the thermal corrections are included. 
At finite temperature, the condition of $m^2<0$ is generalized to
\begin{equation}
    \frac{d^2V(0,T)}{d\phi^2}>0
    ~.
\end{equation}
This requirement should be satisfied at least around the critical temperature.  
For the potential~\eqref{eq:vht1}, the bound on $m^2$ is loosened to  
\begin{equation}
    m^2\lesssim CT_c^2
    ~.
\end{equation}

The critical temperature $T_{c}$ at which the local minimum of the potential at the true vacuum $\phi\neq 0$ is degenerate with that at the 
false vacuum $\phi=0$ is
\begin{equation}
    \label{eq:vhttc2}
    T_{c}^{2}=\frac{\Lambda^{4} m_{\phi}^{4}+2 \Lambda^{2} m_{\phi}^{2} f^{4}-3 f^{8}}{16C\Lambda^{2} f^{4}}
    ~.
\end{equation}
The VEV at the critical temperature is given by
\begin{equation}
    \label{eq:vhtfc2}
    f_{c}^{2}=\frac{3}{2} f^{2}-\frac{m_{\phi}^{2} \Lambda^{2}}{2 f^{2}}
    ~.
\end{equation}
Obviously, both $T_c^2$ and $f_c^2$ are required to be positive to trigger a first-order phase transition, 
giving the bounds~\cite{Grojean2005PRD}
\begin{equation}
    \label{eq:bound2}
    \max \left(\frac{1}{\kappa_m}, \frac{\sqrt{3}}{\sqrt{\kappa_m^{2}+4\varepsilon /3}}\right)<\frac{1}{\kappa_l}< \frac{\sqrt{3}}{\kappa_m}
    ~.
\end{equation}

In order to have a correct direction of the phase transition, the symmetry broken minimum should decrease faster than the symmetric one as the temperature keeps dropping. 
This condition can be expressed as $dV_{\rm HT}/dT^2>0$~\cite{Chiang2020JCAP}, which can be satisfied when $C>0$.  Combining with the bound~\eqref{eq:bound2}, we obtain the resrtriction $\varepsilon >0$.  If we require the tighter upper bound~\eqref{eq:bound1} on $\kappa_m$, then the value of $\varepsilon $ is constrained to be in the range
\begin{equation}
    \varepsilon>\frac{9}{16}\kappa_l^2
    ~.
\end{equation}
One should also ensure the symmetry broken vacuum to be the global minimum at zero temperature, {\it i.e.,}
$V_{\rm HT}(f, 0) < V_{\rm HT}(0, 0)$, which gives the constraint
\begin{equation}
    \label{eq:bound3}
    \kappa_l<\kappa_m
    ~.
\end{equation}
This constraint is valid even when the Coleman-Weinberg potential and the thermal contributions are both taken into account.  Combining with the bound~\eqref{eq:bound1}, we find that $\kappa_l\lesssim \kappa_m$ is required to trigger a first-order phase transition.

The high temperature approximation can be improved by including higher order thermal corrections~\cite{Bodeker2005JHEP}. The corresponding potential is given by
\begin{equation}
    \label{eq:vht2}
    V_{\rm HT}(\phi,T)= -\frac{1}{2}m^2\phi^2+\frac{1}{2}CT^2\phi^2-ET\phi^3+\frac{1}{4}\lambda \phi^4 
    +\frac{1}{\Lambda^2}\left( T^4\phi^2+2T^2\phi^4+\phi^6 \right ),
\end{equation}
where $E=\varepsilon^{3/2}/(4\pi)$.  The high temperature expansion~\eqref{eq:vht1} can approximate the one-loop effective potential~\eqref{eq:Veff} quite well when there exists a tree-level
barrier for separating the two vacua~\cite{Espinosa2012NPB,Chiang2020JCAP}. We further confirm this conclusion for the improved high temperature approximation~\eqref{eq:vht2} by numerically comparing with the effective potential~\eqref{eq:Veff} using various choices of parameters.

\subsection{Parameter scan}

\begin{figure}
    \centering
    \includegraphics[width=100mm,angle=0]{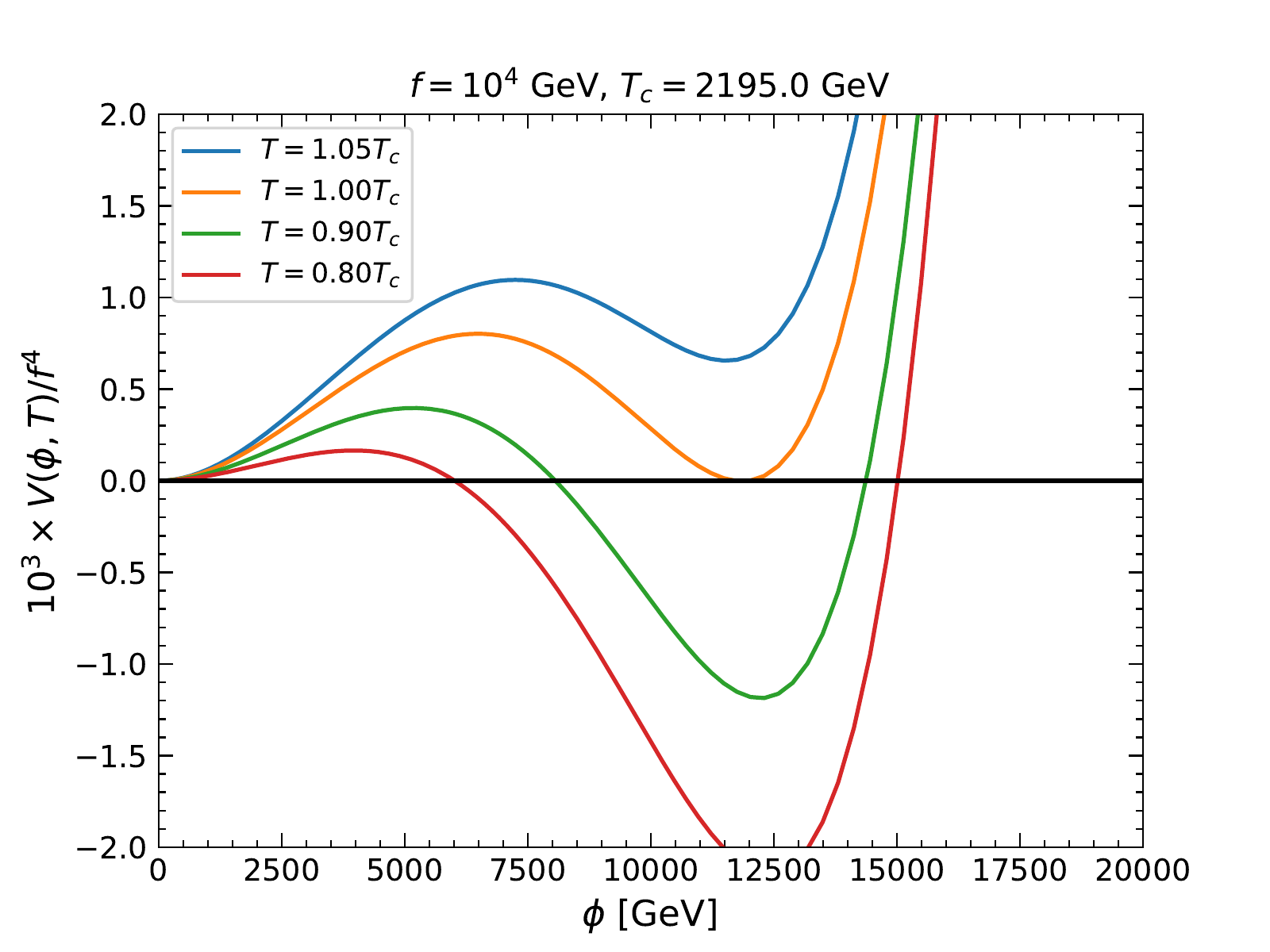}
    \caption{Evolution of the effective potential with temperature, taking $f=10^4$~GeV,
    $\kappa_m=0.15$, $\kappa_l=0.10$, and $\varepsilon=0.51$.}
    \label{fig:potential}
\end{figure}

The phase transition would be first-order if there exists a sufficiently high and wide potential barrier separating the two degenerate vacua of the thermal effective potential at the critical temperature.  As shown in section~\ref{sec:highT}, adding a dimension-6 operator could make the self-coupling $\lambda$ negative. The tree-level barrier could then arise if the parameter $m^2$ is negative (or $d^2V(0,T)/d\phi^2>0$) at the same time. The gear fields can make contributions to the barrier of the effective potential at loop level.  This is given by the third term on the right hand side of the improved high temperature expansion~\eqref{eq:vht2}.  In Fig.~\ref{fig:potential}, we show the evolution of the potential with temperature, using the parameters $f=10^4$~GeV, $\kappa_m=0.15$, $\kappa_l=0.10$, and $\varepsilon=0.51$.  As shown in the figure, the two vacua become degenerate at the critical temperature $T_c=2.195$~TeV and there is a potential barrier between the two vacua.

In search of parameter space that permits a first-order phase transition, we take $f=10^4$~GeV and
make a random scan of the parameters in the following ranges:
\begin{equation}
    10^{-3}\leq\kappa_m\leq 1
    ~,~~
    10^{-3}\leq\kappa_l\leq 1
    ~,~{\rm and}~~
    10^{-3}\leq \varepsilon\leq 1
    ~.
\end{equation}
The upper value of $\kappa_l$ is required by $f\leq \Lambda$ and the upper limit on $\varepsilon$ is to ensure the perturbativity of theory.
Given a set of parameters, we first check various constraints discussed above.  We start from an initial temperature given by eq.~\eqref{eq:vhttc2} and 
find the local VEV minimum around the value given by eq.~\eqref{eq:vhtfc2}.
If the local minimum at the symmetric phase $\phi(T)=0$ is found to be larger (smaller) than the one at the broken phase $\phi(T)\neq 0$, the temperature is increased (decreased) in the next trial.  The critical temperature is then determined by the degenerate condition {\it i.e.,} $V(0,T_c)=V(\phi_c,T_c)$. We find that for most of the sample points, the VEV at critical temperature is larger than that given by eq.~\eqref{eq:vhtfc2}.  We generate one million random floats uniformly for each of the input parameters, among which about 4.8\% are found to be able to trigger afirst-order phase transition.

\begin{figure}
    \centering
    \includegraphics[width=110mm,angle=0]{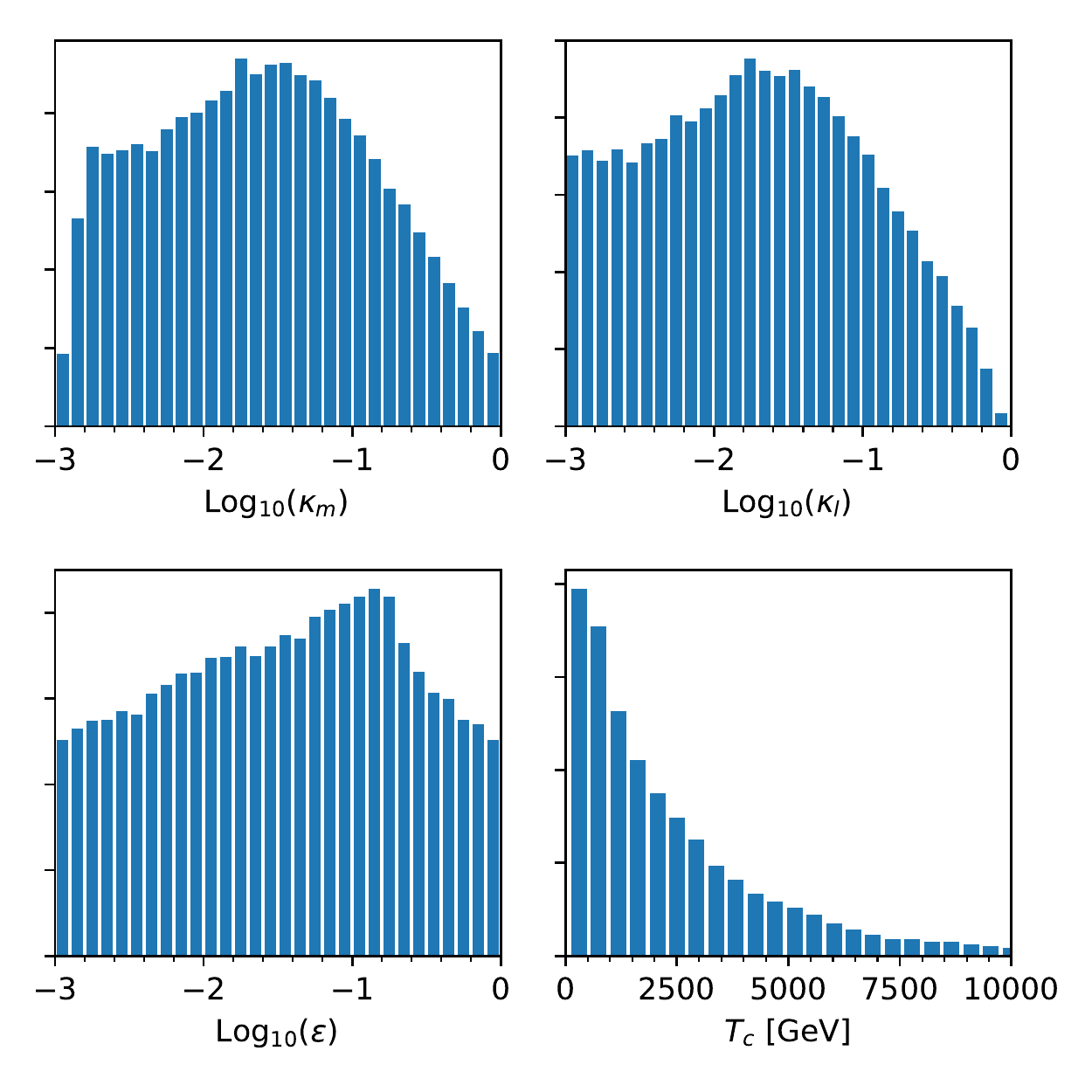}
    \caption{Distributions of the parameters that can trigger a first-order phase transition and the critical temperature distribution.  
    The symmetry breaking scale is fixed at $f=10^4$~GeV.}
    \label{fig:dist_tc}
\end{figure}
\begin{figure}
    \centering
    \includegraphics[width=75mm,angle=0]{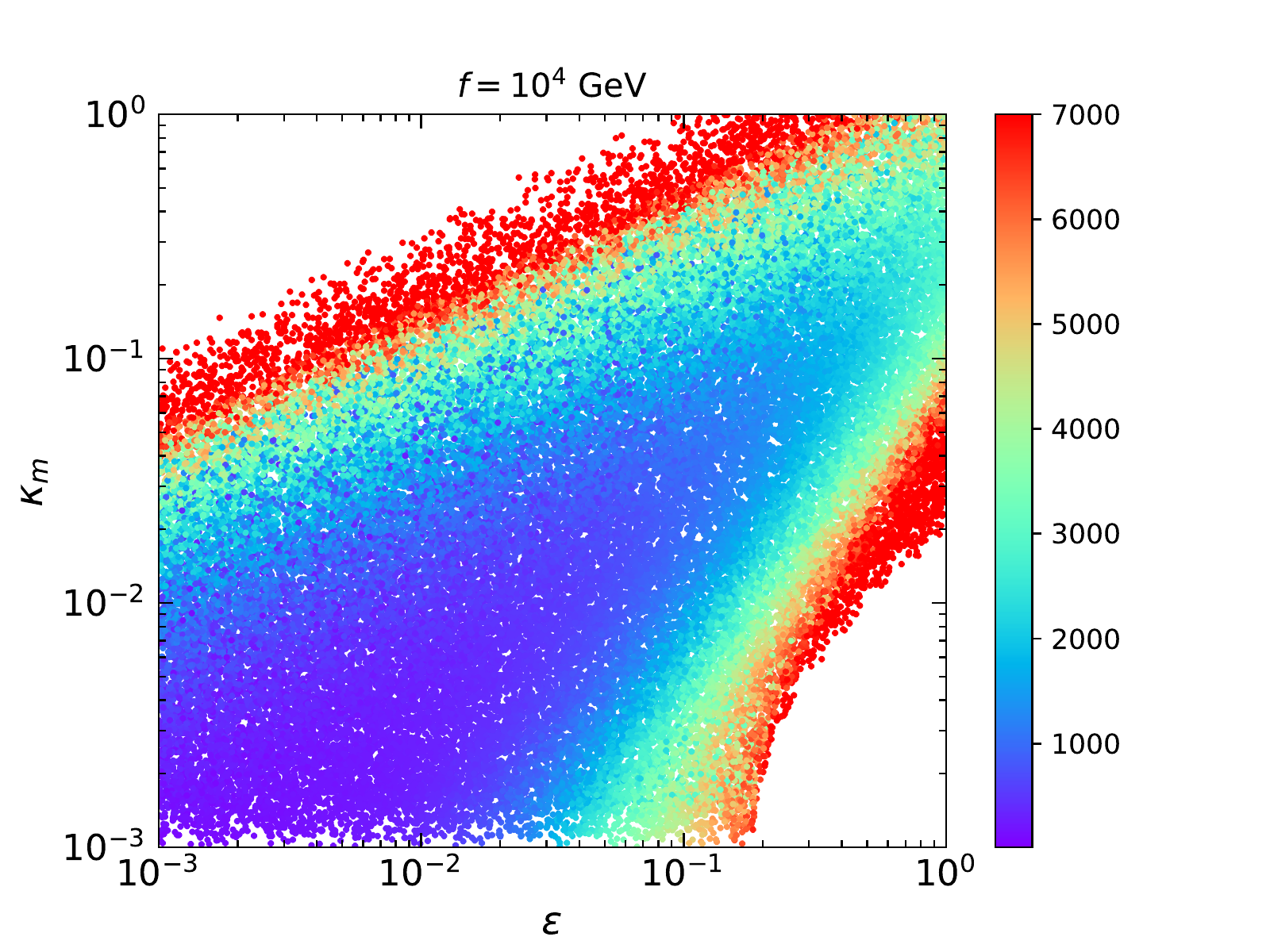}
    \includegraphics[width=75mm,angle=0]{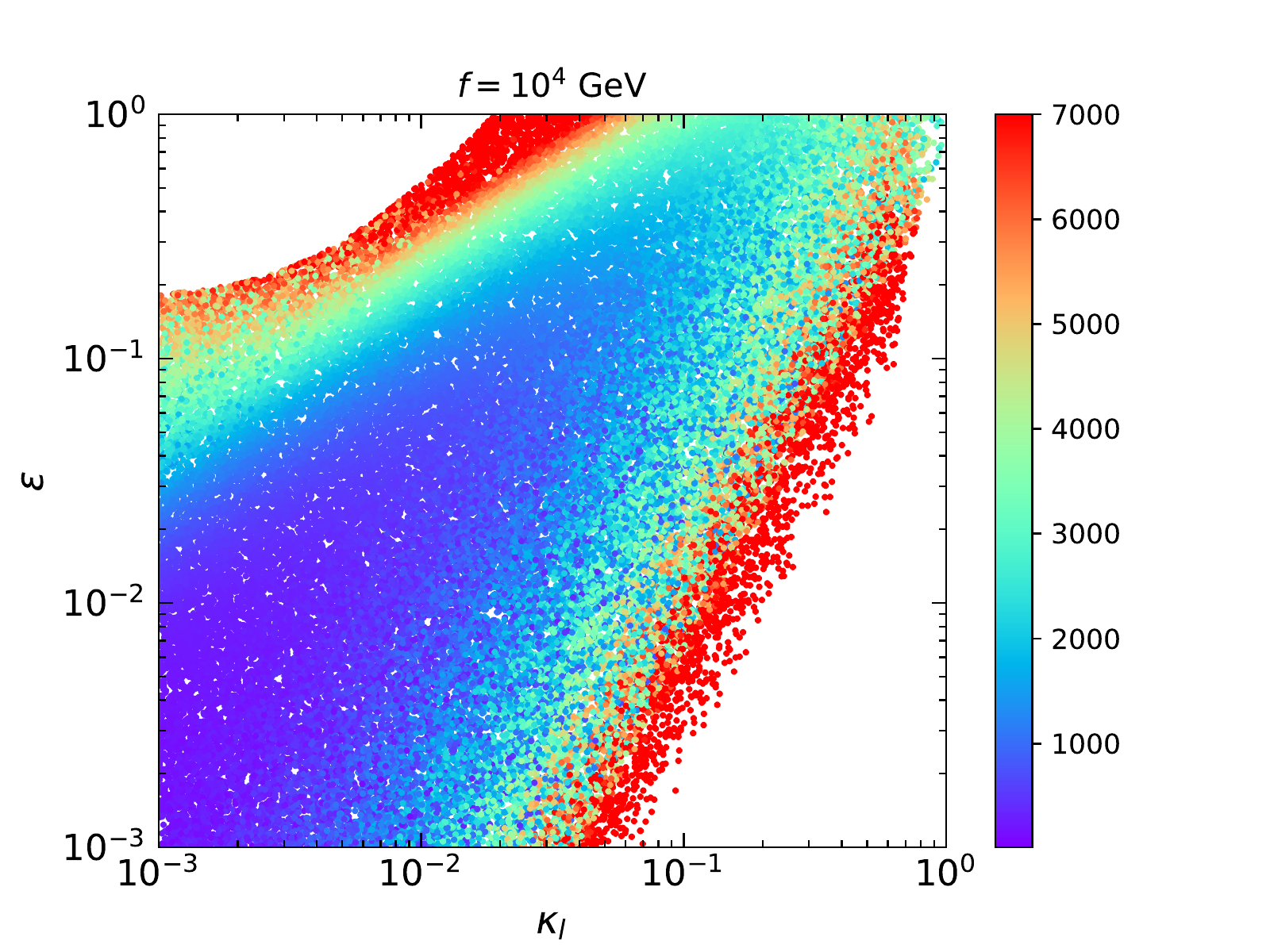}
    \caption{Scatter plots of parameter distributions in the $\varepsilon-\kappa_m$ and $\kappa_l-\varepsilon$ planes.  The color represents the critical temperature in units of GeV.}
    \label{fig:dist_tc2}
\end{figure}

Fig.~\ref{fig:dist_tc} shows the distributions of the parameters and the critical temperature. We have fixed the parameter $f=10^4$~GeV.  For other choices of $f$, the distributions are found to be nearly the same.  We observe that the distribution profile of $\kappa_m$ is similar to that of $\kappa_l$.  This is because of the requirement $\kappa_l\lesssim \kappa_m$ from the above discussions.  
The parameter $\varepsilon$ mainly peaks around 0.1.  The distribution of the critical temperature concentrates in $\lesssim 2.5$~TeV and decreases quickly with the temperature.
Finally, we plot the scatter distributions on the $\varepsilon-\kappa_m$ and $\kappa_l-\varepsilon$ plane in Fig.~\ref{fig:dist_tc2}, from which we can directly observe the variation of the critical temperature with respect to the parameters.

We note that the existence of two degenerate vacua at the critical temperature does not guarantee that a first-order phase transition could successfully happen.  To achieve a successful first-order phase transition, the bubble nucleation of the true vacuum should be triggered successfully as the temperature of the Universe
drops from $T_c$ to a certain value.  This will further constrain the parameter space for the phase transition, as will be discussed in the next section.

\section{Bubble nucleation}
\label{sec:bn}

At high temperatures the Universe is in the symmetric phase, where the vacuum is located at the origin of the scalar field.
As the temperature decreases with the expansion of the Universe, the other minimum of the effective potential appears and becomes the global
minimum when the temperature goes lower than critical temperature.
For a first-order phase transition, the symmetric and broken vacua are separated by a potential barrier.
The tunneling from the metastable minimum to the stable one can proceed through the help of thermal fluctuations. The tunneling process leads to
the decay of the false vacuum and the nucleation of the true vacuum.
The tunneling rate per unit volume and time element is approximately given by~\cite{Apreda2002NPB, Espinosa2008PRD}
\begin{eqnarray}
    \Gamma(T)=A(T)e^{-S_{3}/T}
    ~,
\end{eqnarray}
where $A(T)\simeq [S_3/(2\pi T)]^{3/2}T^{4}$ and $S_3$ denotes the three-dimensional on-shell Euclidean action of instanton.
The probability of bubble nucleations per Hubble volume is defined as
\begin{eqnarray}
    p(T)=\int _{T}^{T_{\rm c}}\frac{\Gamma(x)}{H^{4}(x)}\frac{dx}{x}\approx \left( \frac{T}{H} \right)^4e^{-S_{3}/T}
    ~.
\end{eqnarray}
In a radiation dominated Universe, the Hubble parameter is given by 
\begin{equation}
    H=1.66 g_{*}^{1 / 2} T^{2} / M_{\mathrm{pl}}
    ~,
\end{equation}
where $g_{\ast}\simeq 110$ and $M_{\rm pl}=1.22\times 10^{19}$~GeV is the Planck mass.
The potential barrier decreases with the decrease of the temperature, which improves the probability of the vacuum tunneling.  
The nucleation temperature $T_{\rm n}$ is defined to be one at which the probability of nucleating one bubble per horizon volume 
is of order one, i.e., $p(T)\sim 1$, which can be translated into the following criterion for determining the nucleation 
temperature~\cite{Espinosa2008PRD}
\begin{eqnarray}
  \label{eq:bnc}
  \frac{S_3(T_{\rm n})}{T_{\rm n}}\simeq 4\ln\left( \frac{T_{\rm n}}{H} \right)
  ~.
\end{eqnarray}

For a successful bubble nucleation, the tunneling rate from the flase vacuum to the true vacuum should be large enough to overcome the expansion 
rate of the Universe. It is this criterion that determines whether the first-order cosmological phase transition has successfully proceeded or not. We see that, for one thing, a first-order phase transition needs a barrier to separate the two vacua. This has been fully explored in section~\ref{sec:pt}.
For another thing, the proceeding of the bubble nucleation may be hindered if the barrier is too high or the decrease of the barrier with temperature is too slow. In the following, we will determine the parameter space that satisfies the bubble nucleation criterion.

To determine the bubble nucleation, we have to first obtain the Euclidean action of the $O(3)$ symmetric field configuration $S_3(T)$, which can be written as 
\begin{eqnarray}
    S_{3}(T)=4\pi\int dr~r^2\left [ \frac{1}{2}\left ( \frac{d\phi}{dr} \right )^2+V_{\rm eff}(\phi,T) \right ]
    ~.
\end{eqnarray}
By extremizing the Euclidean action, we obtain the following differential equation
\begin{eqnarray}\label{ce}
    \frac{d^2\phi}{dr^2}+\frac{2}{r}\frac{d\phi}{dr}-\frac{dV}{d\phi}=0
    ~,
\end{eqnarray}
with the boundary conditions
\begin{equation}
    \left.\frac{d \phi}{dr}\right|_{r=0}=0, \quad \lim _{r \rightarrow \infty} \phi(r)=\phi_{\mathrm{false}} \equiv 0
    ~.
\end{equation}
The equation of motion, eq.~(\ref{ce}), can be solved by the traditional overshooting/undershooting method~\cite{Apreda2002NPB}.
In this work, we employ the $\textsf{CosmoTransitions~2.0.2}$ package~\cite{Wainwright2012PLB} to perform the numerical calculations of 
the bubble profile and Euclidean action. Afterwards, we use eq.~\eqref{eq:bnc} to determine the nucleation temperature $T_{n}$.
The first-order phase transition is usually completed after percolation of the true vacuum bubbles. 
The production of GWs is significant at the percolation time (temperature), at which 34\% of the false vacuum has been converted to the true 
vacuum~\cite{Cai2017JCAP,Ellis2019JCAP,Wang2020JCAP}. In this work, we assume that the percolation takes place soon after the nucleation of the
true vacua, which leads to the commonly used condition $T_*\simeq T_n$, where $T_*$ is the GW generation temperature~\cite{Espinosa2008PRD,Ellis2020JCAP}.

The stochastic GWs generated by the first-order phase transition can be fully characterized by the knowledge of two primary parameters~\cite{Kamionkowski1994PRD}. One of them is the latent heat normalized by the radiation energy density in the plasma
\begin{eqnarray}
    \label{eq:ala}
    \alpha =\frac{\epsilon (T_n)}{\rho _{\rm rad}(T_n)}
    ~,
\end{eqnarray}  
where $\rho_{\rm rad}=\pi^2g_{\ast}T^4/30$ is the radiation energy density in the plasma, and the latent heat associated with the phase transition is given by
\begin{eqnarray}
    \epsilon (T)=T\frac{\partial \Delta V_{bs}(T)}{\partial T}- \Delta  V_{bs}(T)
    ~,
\end{eqnarray}
where $\Delta V_{bs}(T)\equiv V_{\rm eff}(f(T),T)-V_{\rm eff}(0,T)$ is the potential difference between the broken phase and the symmetric phase at temperature $T$.   
The parameter $\alpha$ is related to the maximum available energy budget for gravitational wave emissions.
The other parameter relevant to the GW production is defined as 
\begin{equation}
    \frac{\beta}{H_{n}}=\left.T_{n} \frac{d}{d T}\left(\frac{S_{3}(T)}{T}\right)\right|_{T=T_{n}}
    ~.
\end{equation}
The parameter $\beta$ represents the rate of time variation of the nucleation rate, whose inverse gives the duration of the bubble nucleation. 
Consequently, $\beta/H_n$ defines the characteristic frequency of the GW spectrum produced from the phase transition.

\begin{figure}
    \centering
    \includegraphics[width=75mm,angle=0]{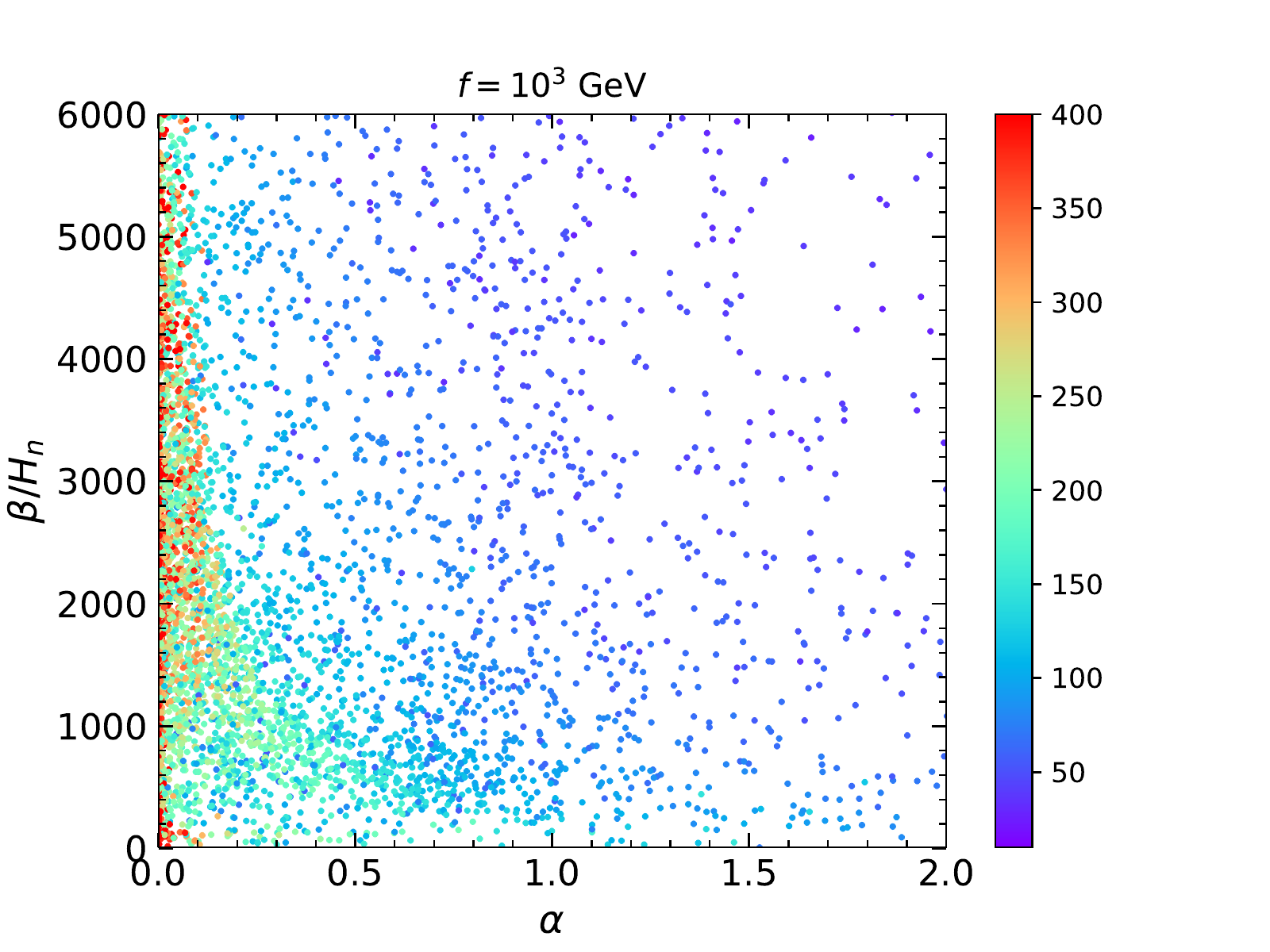}
    \includegraphics[width=75mm,angle=0]{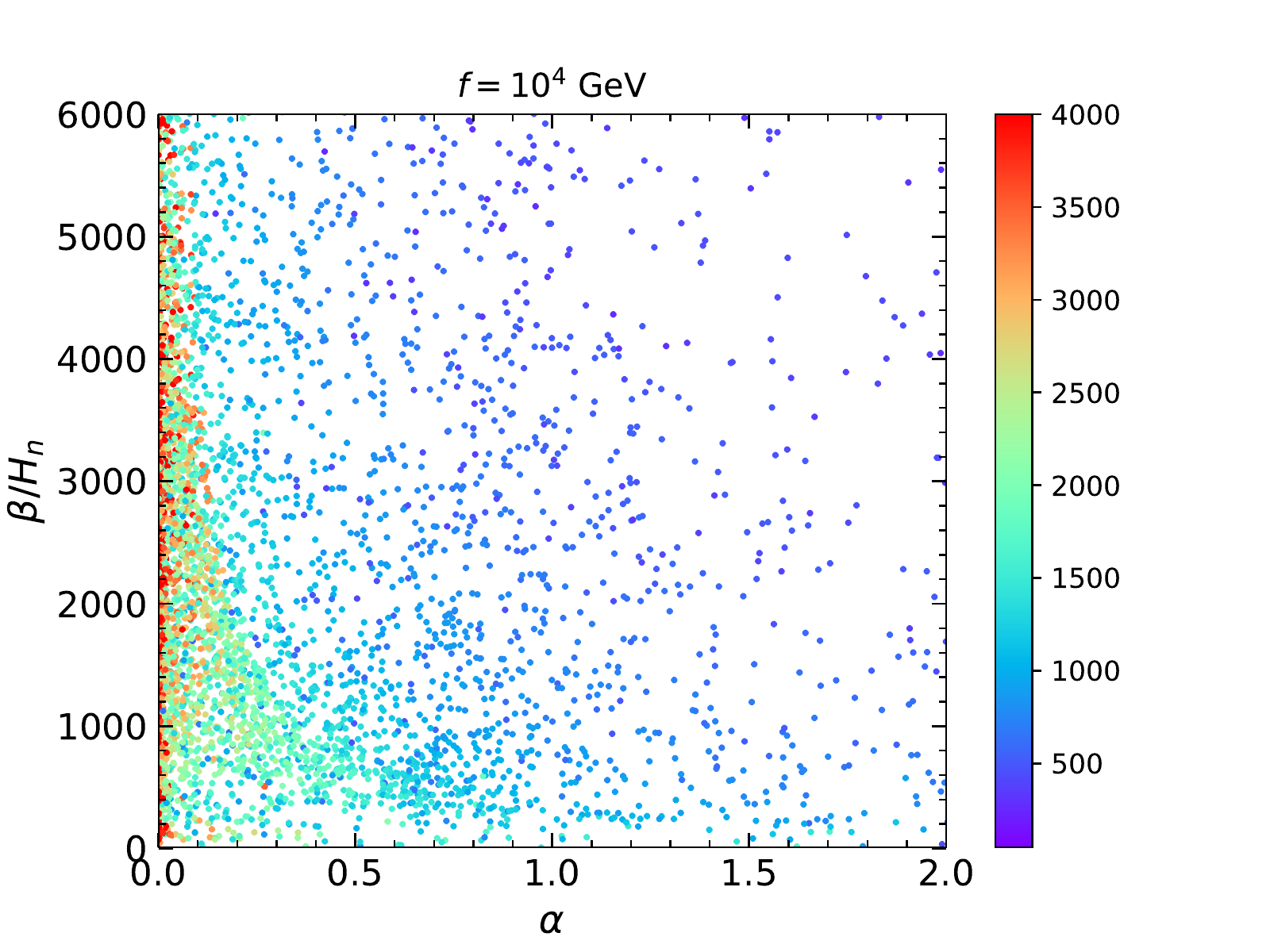}\\
    \includegraphics[width=75mm,angle=0]{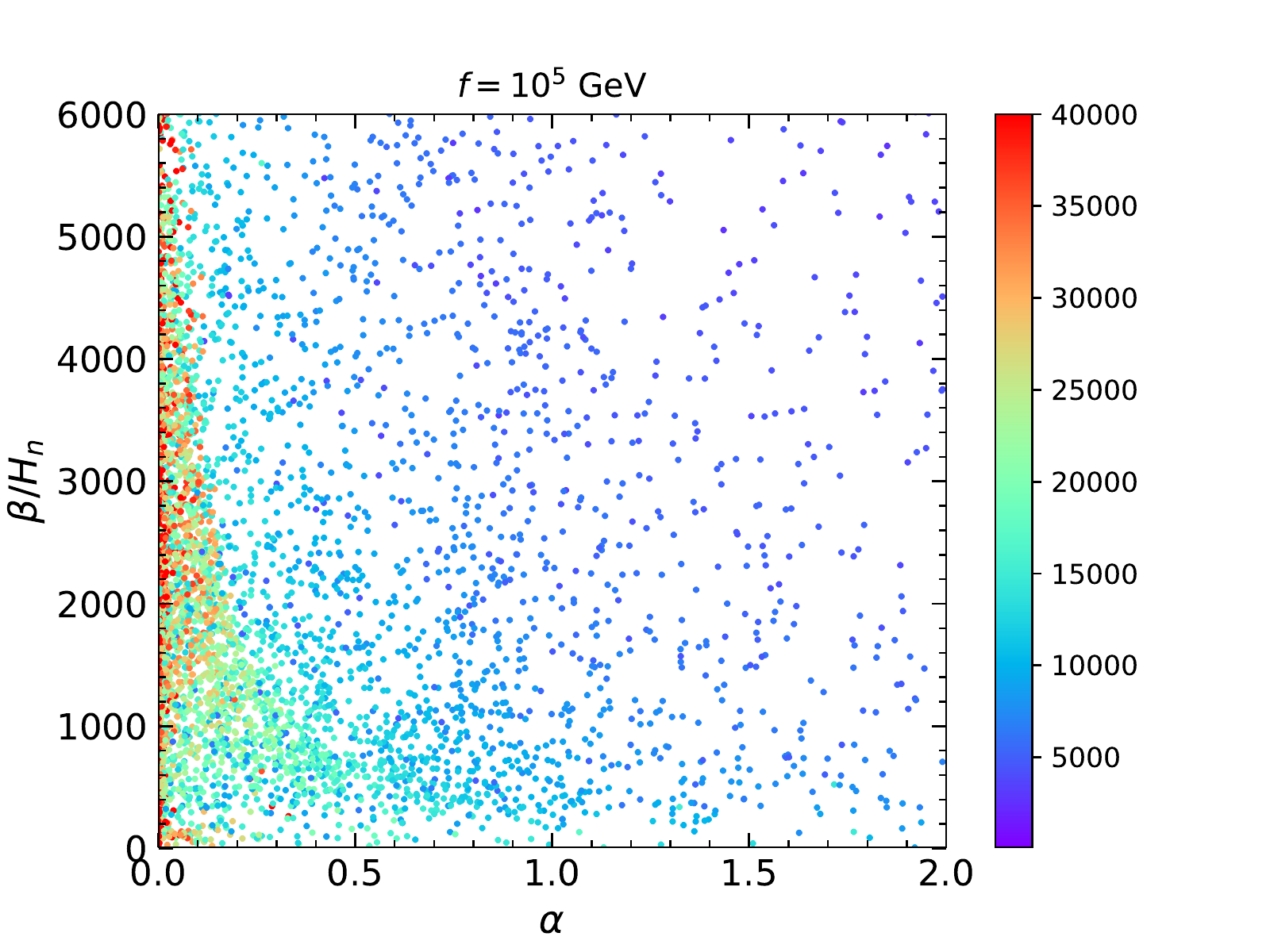}
    \includegraphics[width=75mm,angle=0]{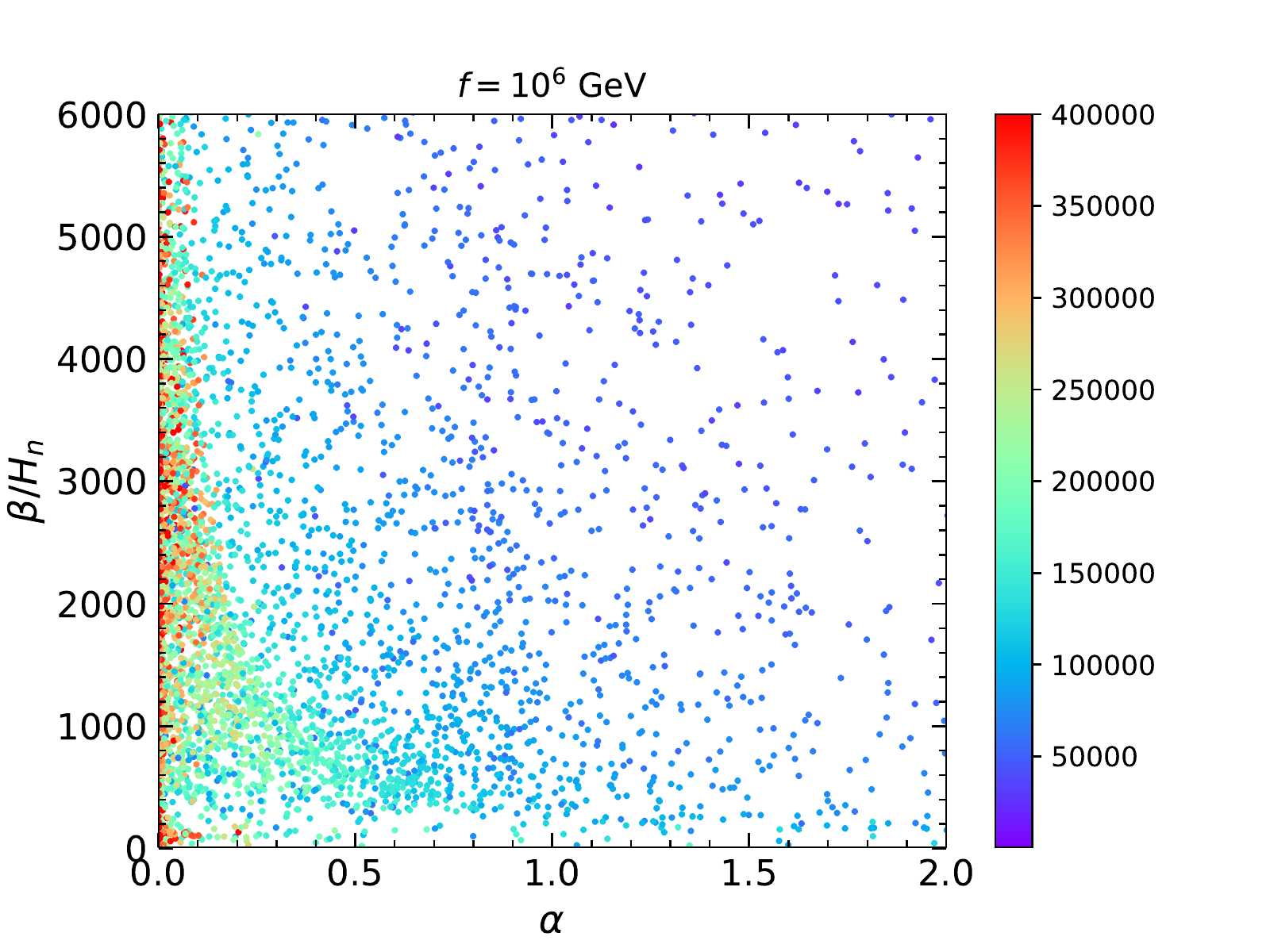}
    \caption{Distributions of the GW parameters $\alpha$ and $\beta/H_n$.  The color bars indicate
    the nucleation temperature. For the upper left, upper right, lower left and lower right plots, the symmetry breaking scales are taken as $f=10^3$~GeV, $10^4$~GeV, $10^5$~GeV, and $10^6$~GeV, respectively.}
    \label{fig:ab}
\end{figure}

\begin{figure}
    \centering
    \includegraphics[width=110mm,angle=0]{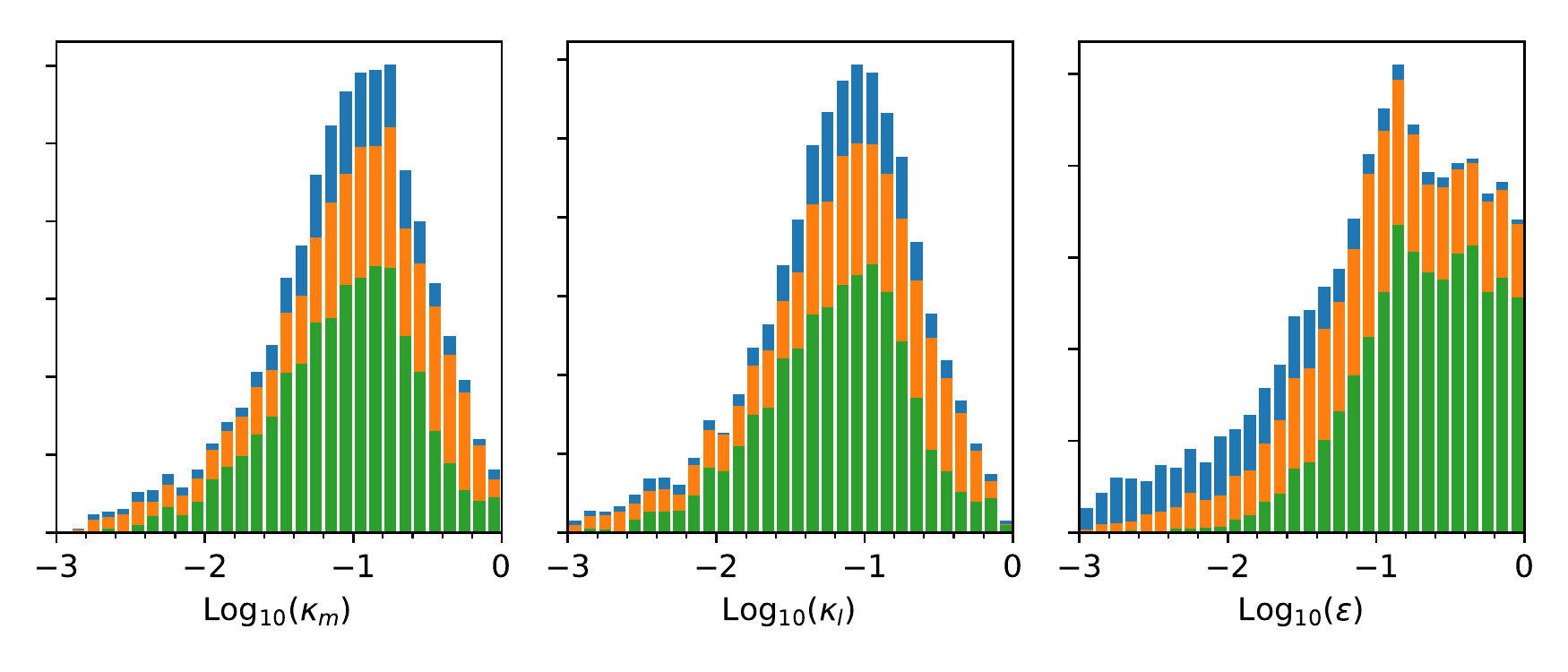}\\
    \includegraphics[width=110mm,angle=0]{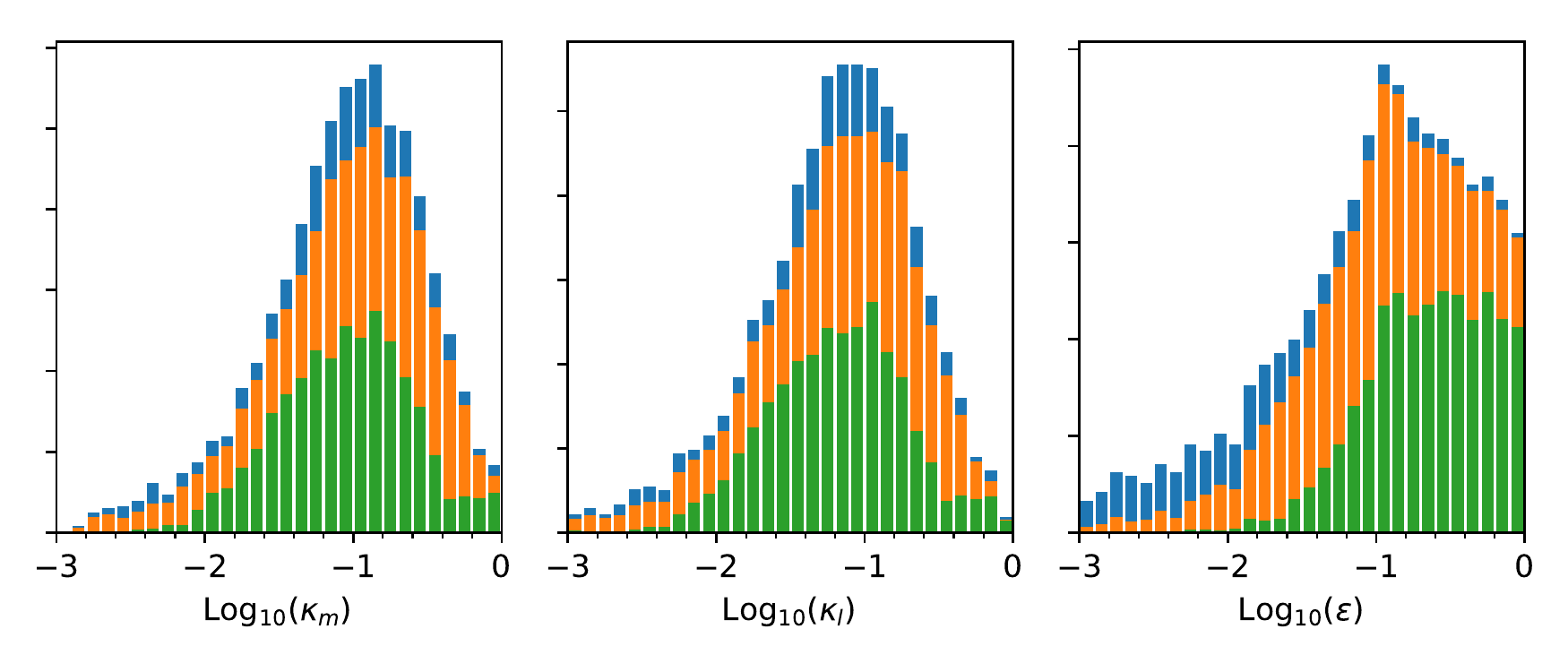}\\
    \includegraphics[width=110mm,angle=0]{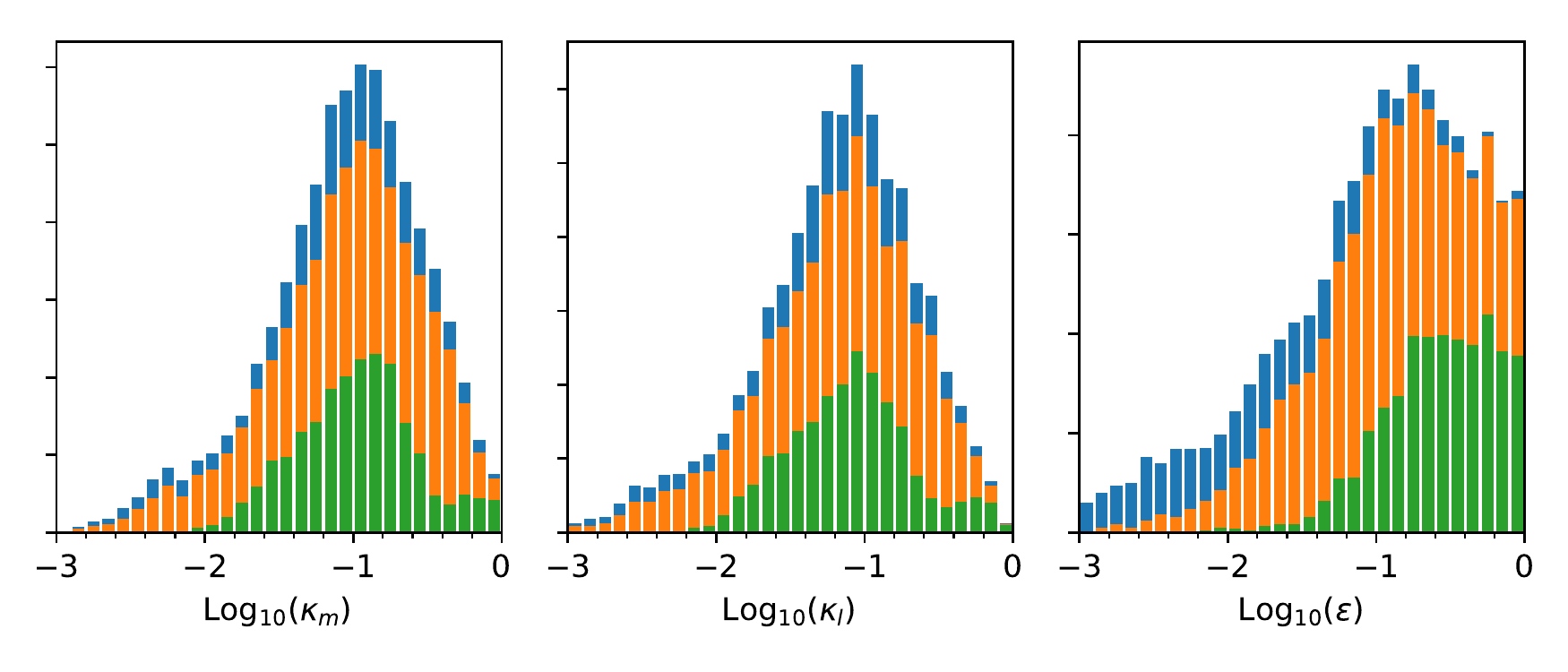}\\
    \includegraphics[width=110mm,angle=0]{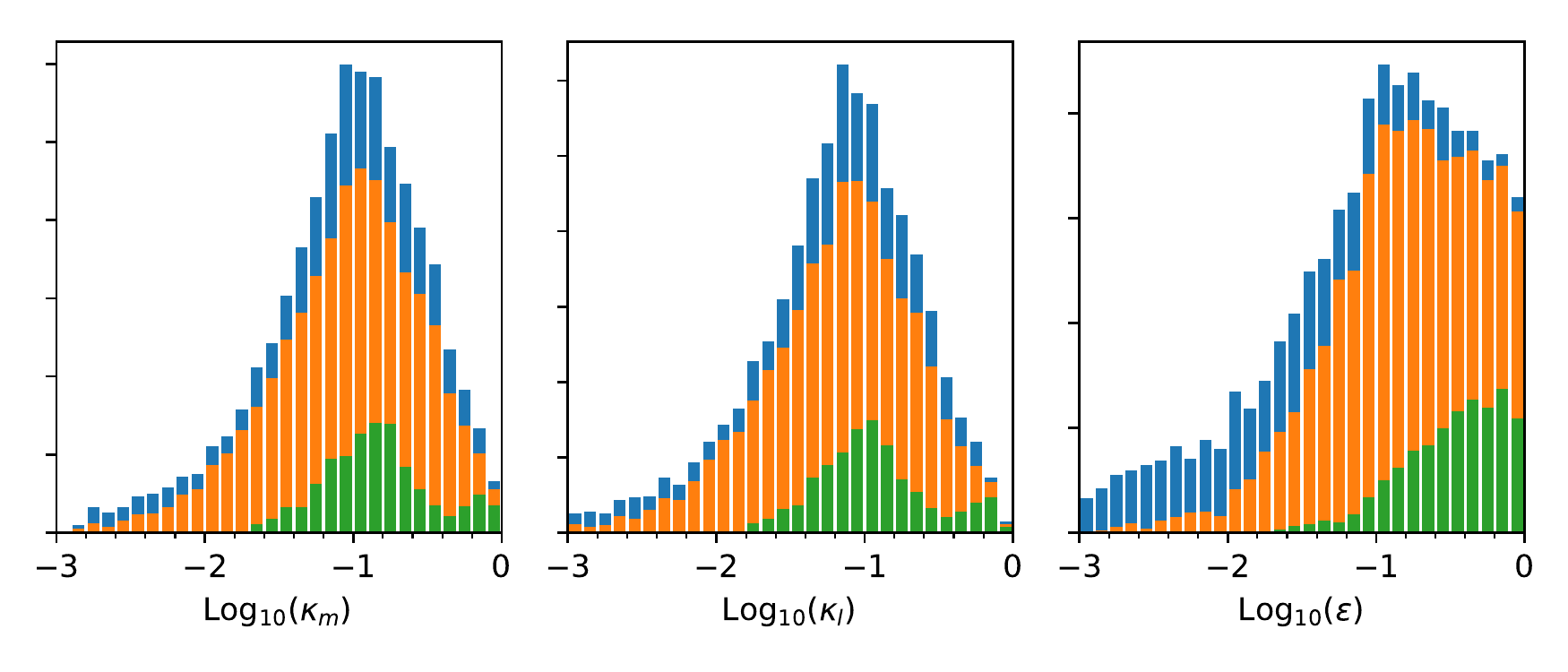}
    \caption{Histograms of the parameters that satisfy the nucleation condition. The blue histograms represent the total samples, 
    and the orange and green histograms denote those samples that are detectable by the BBO and LISA interferometers, respectively. 
    From top to bottom, the symmetry breaking scales are fixed at $f=10^3$~GeV, $10^4$~GeV, $10^5$~GeV, and $10^6$~GeV. 
    }
    \label{fig:dist_tn}
\end{figure}

In the scatter plots of Fig.~\ref{fig:ab}, we show the calculated results of $\alpha$ and $\beta/H_n$, with the the corresponding nucleation temperature $T_n$ indicated by the colored dots. We find about 15\% of the sample points that trigger a first-order phase transition 
(see Fig.~\ref{fig:dist_tc}) can also satisfy the bubble nucleation condition. The plot shows that the nucleation temperature tends to be lower for larger $\alpha$, in agreement with the observation that the GWs can be stronger when they are produced at a lower nucleation temperature~\cite{Ellis2019JCAP,Ellis2020JCAP,Ellis2020}.
The distributions of parameters $\kappa_m$, $\kappa_l$, and $\varepsilon$ are shown by the blue histograms in Fig.~\ref{fig:dist_tn}.
From the upper plots to the lower plots, the $U(1)$ symmetry breaking scale $f$ is taken as $10^3$~GeV, $10^4$~GeV, $10^5$~GeV, and $10^6$~GeV, respectively. 
Compared with the parameter distributions for generating a first-order phase transition, as shown in Fig.~\ref{fig:dist_tc}, 
$\kappa_m$, $\kappa_l$, and $\varepsilon$ have been restricted to be $\gtrsim 10^{-2}$ by the criterion of the successful bubble nucleation.

\section{Gravitational waves from phase transition}
\label{sec:GWPT}

\subsection{Gravitational wave sources}

We first briefly review the numerical simulation results of the GW spectra produced from a first-order phase transition.
During the percolation of the bubbles, there exist three processes that can produce the GWs~\cite{Espinosa2010JCAP,No2011PRD,Caprini2016JCAP}:

\begin{itemize}
\item[1.] {\bf Bubble Collisions}. GWs produced from this process depends only on the dynamics of the scalar field.
The GW spectrum from the collisions of bubble walls estimated by the numerical simulations~\cite{Huber2008JCAP} is given by
\begin{equation}
  \label{eq:gwspt1}
  h^{2} \Omega_{\mathrm{\rm col}}(f)=1.67 \times 10^{-5}\left(\frac{H_{n}}{\beta}\right)^{2}\left(\frac{\kappa_{\rm col}\alpha}
  {1+\alpha}\right)^{2}\left(\frac{100}{g_{*}}\right)^{\frac{1}{3}}\left(\frac{0.11 v_{w}^{3}}{0.42+v_{w}^{2}}\right) 
  \frac{3.8\left(f / f_{\mathrm{col}}\right)^{2.8}}{1+2.8\left(f / f_{\mathrm{col}}\right)^{3.8}}
  ~,
\end{equation}
where the red-shifted peak frequency of the GW spectrum from bubble collision is
\begin{equation}
    f_{\mathrm{col}}=16.5 \times 10^{-3} \mathrm{mHz}\left(\frac{0.62}{1.8-0.1 v_{w}+v_{w}^{2}}\right)
    \left(\frac{\beta}{H_{n}}\right)\left(\frac{T_{n}}{100 \mathrm{GeV}}\right)\left(\frac{g_{*}}{100}\right)^{\frac{1}{6}}
    ~.
\end{equation}
The efficiency factors $\kappa_{\rm col}$ indicates the fraction of latent heat that is transformed into the kinetic energy of bubbles. 
In the case of non-runaway bubbles, the bubble walls will reach a terminal velocity and the latent energy transferred into the scalar field is 
negligible. For the runaway bubbles, however, most of the bubble energy is dissipated into the surrounding plasma and very little energy 
is deposited in the bubble walls~\cite{Bodeker2017JCAP}. 
Both cases lead to a negligible GW spectrum from the bubble collisions, and thus, in this work we do not take into account 
the bubble collision's contributions.
In the following, we will restrict ourselfs to the case of non-runaway bubbles, in which the GWs can be effectively produced by the 
sound waves and turbulence. 

\item[2.] {\bf Sound waves.} They are generated subsequently after the bubble collisions. 
Numerical simulations indicate that the durations of sound waves and turbulence as active sources of GWs are typically much longer 
than the collisions of the bubble walls. 
The GW spectrum generated by sound waves propagating in the plasma is approximated by~\cite{Hindmarsh2015PRD}
\begin{equation}
  \label{eq:gwspt2}
  h^{2} \Omega_{\mathrm{sw}}(f)=2.65 \times 10^{-6}\left(\frac{H_{*}}{\beta}\right)\left(\frac{\kappa_{\rm sw} \alpha}
  {1+\alpha}\right)^{2}\left(\frac{100}{g_{*}}\right)^{\frac{1}{3}} v_{w} 
  \left(\frac{f}{f_{\rm{sw}}}\right)^{3}\left(\frac{7}{4+3\left(f / f_{\mathrm{sw}}\right)^{2}}\right)^{7 / 2},
\end{equation}
where the red-shifted peak frequency of the GW spectrum from sound waves is
\begin{equation}
    f_{\mathrm{sw}}=1.9 \times 10^{-2} \mathrm{mHz} \frac{1}{v_{w}}\left(\frac{\beta}{H_{n}}\right)
    \left(\frac{T_{n}}{100 \mathrm{GeV}}\right)\left(\frac{g_{*}}{100}\right)^{\frac{1}{6}}.
\end{equation}
The efficiency factors $\kappa_{\rm sw}$ indicates the fraction of latent heat that is transformed into the bulk motion of the plasma. 
For the non-runaway bubbles, the efficiency factor for the sound wave contribution is then given by
\begin{equation}
    \label{eq:ksw}
    \kappa_{\mathrm{sw}}\simeq \frac{\alpha}{0.73+0.083 \sqrt{\alpha}+\alpha}
    ~,~~{\rm for}~~v_w\sim 1.0
    ~.
\end{equation}

\item[3.] {\bf Turbulence.} As the sound waves, turbulence in the plasma forms after the bubble collisions. 
Simulations show that only a small fraction $\delta \sim 5-10\%$ of the bulk motion from the bubble walls is converted into turbulence. 
Since the GWs from sound waves decay much faster, the GWs from turbulence could play a dominant role at high frequencies.  
The modeling of turbulence is far from settled.  In this work, we adopt GW spectrum from turbulence as follows~\cite{Caprini2009JCAP}
\begin{equation}
  \label{eq:gwspt3}
  h^{2} \Omega_{\mathrm{turb}}(f)=3.35 \times 10^{-4}\left(\frac{H_{*}}{\beta}\right)\left(\frac{\kappa_{\mathrm{turb}} \alpha}
  {1+\alpha}\right)^{\frac{3}{2}}\left(\frac{100}{g_{*}}\right)^{1 / 3} v_{w} 
  \frac{\left(\frac{f}{f_{\mathrm{turb}}}\right)^{3}}{\left[1+\frac{f}{f_{\mathrm{turb}}}\right]^{\frac{11}{3}}\left(1+\frac{8\pi f}{H_{0}}\right)}
  ~,
\end{equation}
where the red-shifted Hubble constant observed today is given by
\begin{equation}
  H_{0}=16.5 \times 10^{-3} \mathrm{mHz}\left(\frac{T_{n}}{100 \mathrm{GeV}}\right)\left(\frac{g_{*}}{100}\right)^{\frac{1}{6}}.
\end{equation}
The red-shifted peak frequency of the GW spectrum from turbulence is
\begin{equation}
    \label{eq:fturb}
    f_{\text {turb }}=2.7 \times 10^{-2} \mathrm{mHz} \frac{1}{v_{w}}\left(\frac{\beta}{H_{n}}\right)
    \left(\frac{T_{n}}{100 \mathrm{GeV}}\right)\left(\frac{g_{*}}{100}\right)^{\frac{1}{6}}.
\end{equation}
The efficiency factor for turbulence $\kappa_{\rm turb}$ is related to $\kappa_{\rm sw}$ by $\kappa_{\rm turb}=\delta\kappa_{\rm sw}$,
where we take $\delta=0.1$ in this work.

\end{itemize}

\begin{figure}
    \centering
    \includegraphics[width=75mm,angle=0]{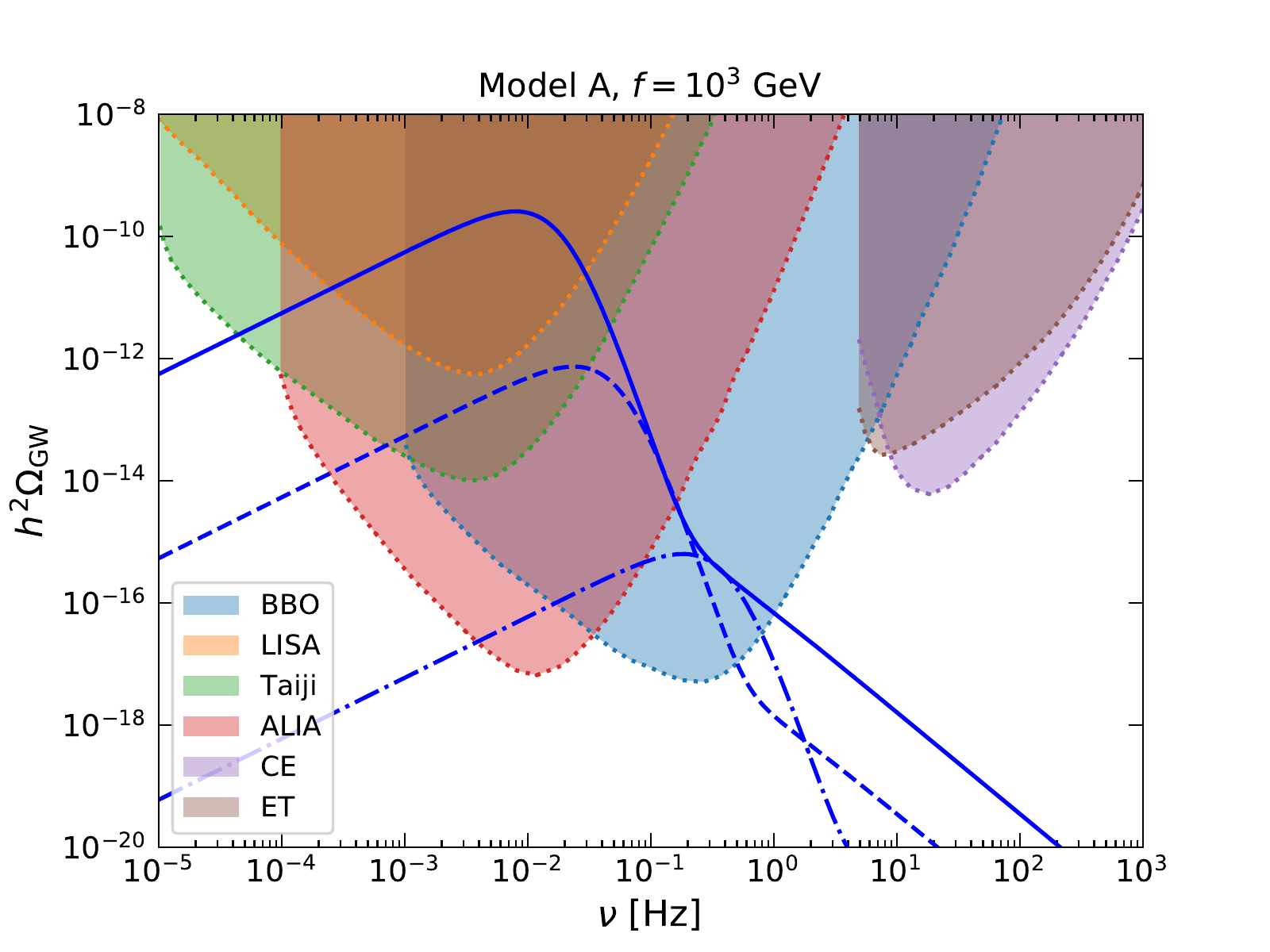}
    \includegraphics[width=75mm,angle=0]{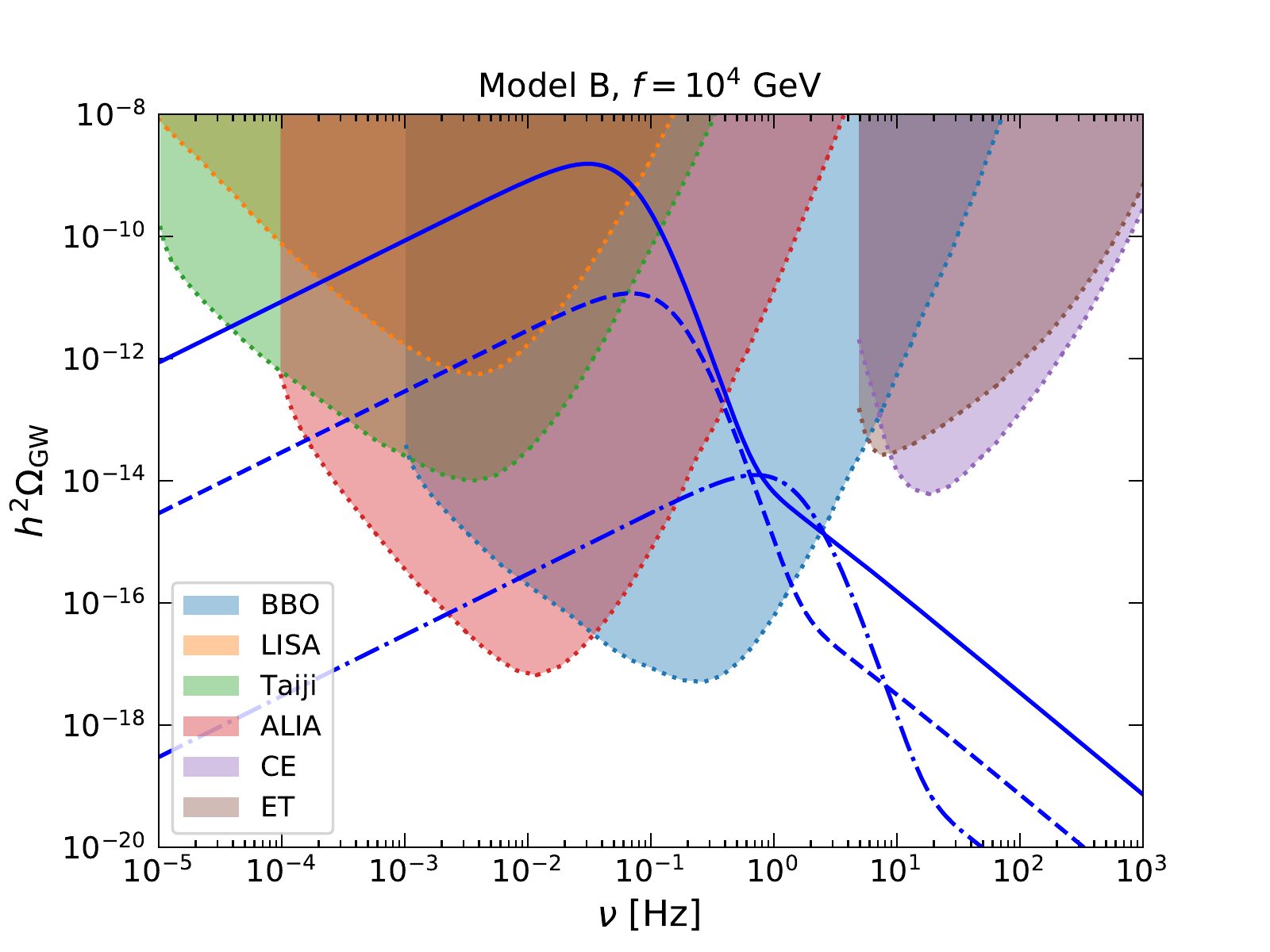}\\
    \includegraphics[width=75mm,angle=0]{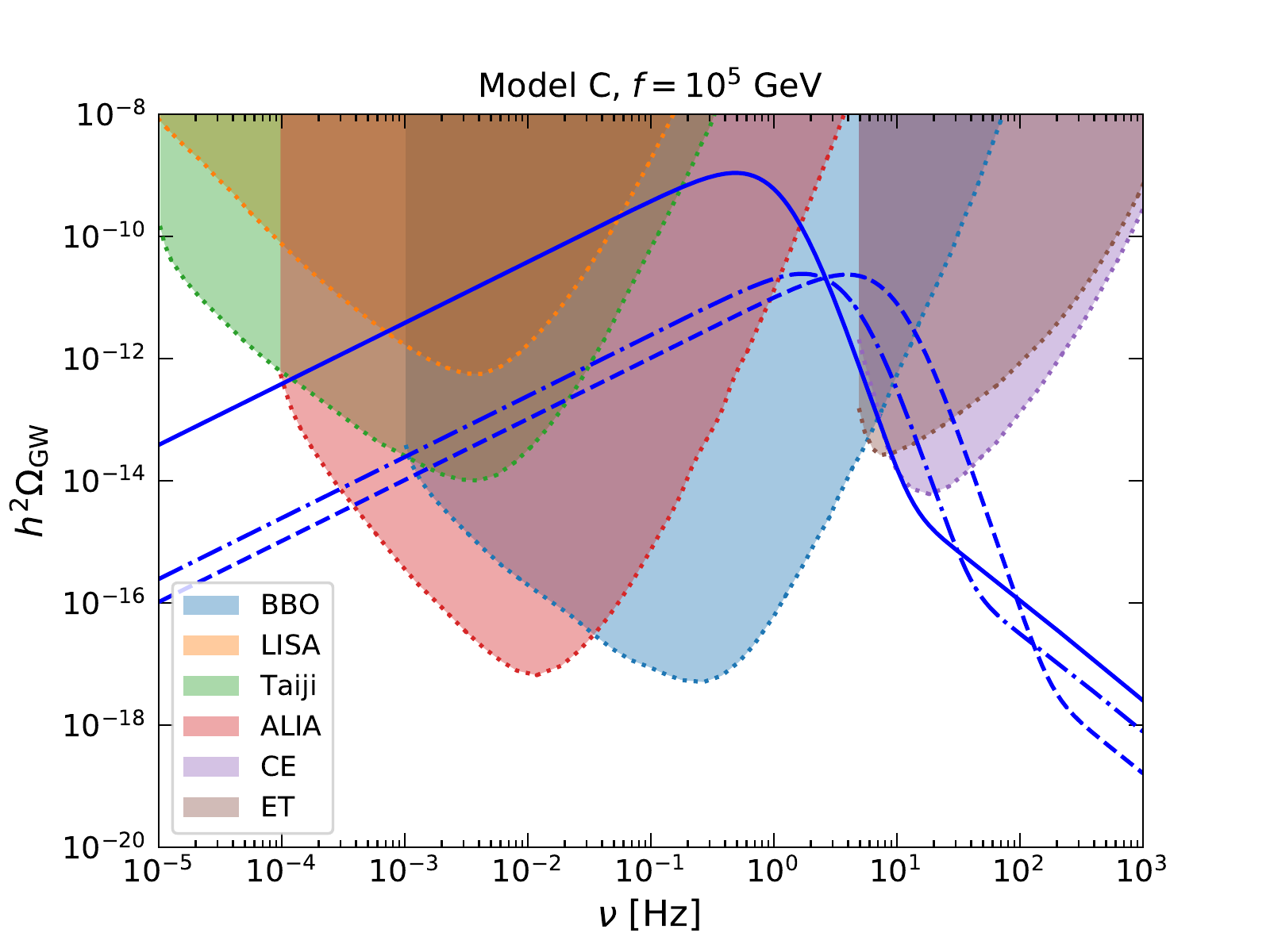}
    \includegraphics[width=75mm,angle=0]{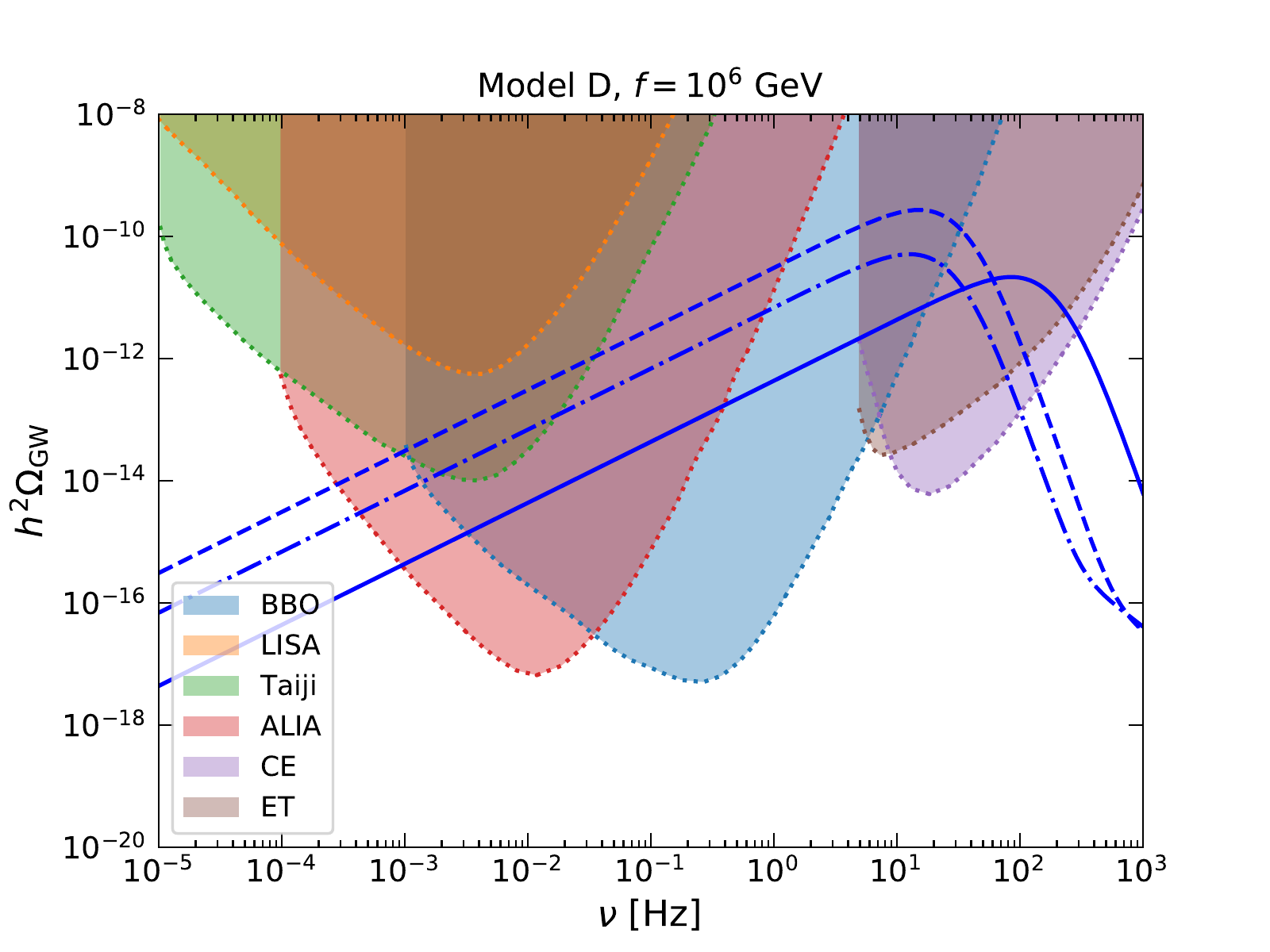}
    \caption{GW spectra from phase transition (curves) and various experimental sensitivities (colored patches) as a function of frequency.
    The solid, dashed, and dash-dotted curves represent the results in model $\rm M_1$, $\rm M_2$, and $\rm M_3$ (where M=A, B, C, and D).
    For the upper left, upper right, lower left and lower right plots, the symmetry breaking scales are taken as $f=10^3$~GeV, $10^4$~GeV, $10^5$~GeV, and $10^6$~GeV, respectively.
    The parameters used in this plots are summarized in Table.~\ref{tab:i}.}
    \label{fig:GWsp}
\end{figure}

The amplitude and peak frequency of the GW spectrum also depend on the bubble wall velocity $v_w$, which is the expanding speed of the true vacuum.
In this work, we take $v_w\simeq 1.0$ for the calculations of GW spectra.
The total stochastic GW spectrum is approximately given by adding up these three contributions:
\begin{equation}
  h^{2} \Omega_{\mathrm{GW}} \simeq h^{2} \Omega_{\rm col}+h^{2} \Omega_{\mathrm{sw}}+h^{2} \Omega_{\mathrm{turb}}.
\end{equation}
Note that as explained above, the bubble collision's contribution has been neglected in our calculation.

\subsection{Gravitational wave detections}

For the experimental investigation of the stochastic GW signals, one often adopts the frequentist approach where the detectability of the signals is measured by the corresponding signal-to-noise ratio (SNR)~\cite{Caprini2016JCAP}
\begin{equation}
  \rho=\sqrt{\mathcal{N}\mathcal{T}_{\rm obs} \int_{f_{\mathrm{min}}}^{f_{\mathrm{max}}} 
  df\left[\frac{h^{2} \Omega_{\mathrm{GW}}(f)}{h^{2} \Omega_{\mathrm{exp}}(f)}\right]^{2}}
  ~,
\end{equation}
where $\mathcal{N}$ is the number of independent observatories of the experiment, $\mathcal{T}_{\rm obs}$ is the duration of the mission, and
$h^2\Omega_{\mathrm{exp}}$ denotes the sensitivity of a GW experiment.
Following ref.~\cite{Caprini2016JCAP}, we take the SNR threshold value $\rho_{\rm thr}=10$, above which the GW signal is detectable for the experiment.

The future space-based GW interferometers, including LISA~\cite{LISA2017,LISA2019CQG}, Taiji~\cite{Hu2017NSR,Ruan2020NA}, ALIA~\cite{ALIA2014JPCS}, DECIGO~\cite{DECIGO2017}, and BBO~\cite{BBO2006CQG} are able to explore the GW signals with the frequencies in the range of $\sim 10^{-5}-10^2$~Hz.  Higher frequencies ($\sim 10-10^3$~Hz) GW signals are expected to be probed by the ground-based GW observatories such as aLIGO~\cite{LIGO2019}, ET~\cite{ET2010CQG}, and CE~\cite{CE2017CQG}.

We show the distributions of parameters with detectibility for BBO and LISA interferometers in Fig.~\ref{fig:dist_tn}.
We use $\mathcal{N}=1$ for the auto-correlated experiment, LISA, while for the cross-correlated experiments, BBO, we take $\mathcal{N}=2$.
Following ref.~\cite{Chiang2020JHEP}, we assume a mission duration of $\mathcal{T}_{\rm obs}=4$ years for both interferometers.
The observation frequency ranges are set in the range of $10^{-3}-10^2$ and $10^{-5}-1$ Hz for BBO and LISA, respectively. 
The experimental sensitivities of BBO and LISA are summarized in appendix E of ref.~\cite{Chiang2020JHEP}.
The orange and green histograms represent the detectable regions for BBO and LISA, respectively.
As indicated in the figure, the BBO experiment can probe almost all of the parameter space that can successfully generate a first-order phase transition.
The GW signals produced at the scale $f=10^{3}$~GeV in the clockwork model can be effectively detected by the LISA experiment.  The nucleation temperature increases with $f$, leading to higher peak frequencies of the GW signals.  As a result, it is difficult to probe the GW signals induced by the $U(1)$ symmetry breaking at a scale of $f=10^{6}$~GeV at the LISA interferometer.
We will show in section~\ref{sec:GWDW} that the domain walls produced after the phase transition at the symmetry breaking scale 
$f\gtrsim 10^{6}$~GeV could dominate the energy density of the Universe, which is not compatible with cosmological observations.
Thus, here we do not further consider those GW signals produced from the phase transition above the scale of $f=10^{6}$~GeV.

\begin{table}[]
    \caption{A summary of models and parameters}
    \begin{tabular}{|c|c|l|l|l|l|l|l|l|}
    \hline
    \multicolumn{1}{|l|}{}                                                            & Models    & $\kappa_m$ & $\kappa_l$ & $\varepsilon$ & $T_c$ {[}GeV{]} & $T_n$ {[}GeV{]} & $\alpha$ & $\beta/H_n$ \\ \hline
    \multirow{3}{*}{\begin{tabular}[c]{@{}c@{}}Model A\\ $f=10^3$ GeV\end{tabular}}   & $\rm A_1$ & 0.176      & 0.128      & 0.548         & 231.4           & 115.4           & 0.715    & 751.4       \\ \cline{2-9} 
                                                                                      & $\rm A_2$ & 0.120      & 0.092      & 0.094         & 207.1           & 126.6           & 0.133    & 2046.6      \\ \cline{2-9} 
                                                                                      & $\rm A_3$ & 0.449      & 0.332      & 0.309         & 417.9           & 383.3           & 0.025    & 5188.6      \\ \hline
    \multirow{3}{*}{\begin{tabular}[c]{@{}c@{}}Model B\\ $f=10^4$ GeV\end{tabular}}   & $\rm B_1$ & 0.228      & 0.181      & 0.440         & 2287.3          & 969.2           & 1.220    & 348.2       \\ \cline{2-9} 
                                                                                      & $\rm B_2$ & 0.080      & 0.053      & 0.511         & 2524.1          & 645.5           & 0.261    & 1163.0      \\ \cline{2-9} 
                                                                                      & $\rm B_3$ & 0.101      & 0.071      & 0.086         & 1979.4          & 1597.2          & 0.054    & 4877.7      \\ \hline
    \multirow{3}{*}{\begin{tabular}[c]{@{}c@{}}Model C\\ $f=10^{5}$ GeV\end{tabular}} & $\rm C_1$ & 0.151      & 0.115      & 0.538         & 21686.2         & 8087.8          & 1.472    & 660.1       \\ \cline{2-9} 
                                                                                      & $\rm C_2$ & 0.029      & 0.019      & 0.092         & 9494.1          & 4607.1          & 0.763    & 9355.4      \\ \cline{2-9} 
                                                                                      & $\rm C_3$ & 0.317      & 0.227      & 0.807         & 29054.8         & 20595.1         & 0.307    & 907.8       \\ \hline
    \multirow{3}{*}{\begin{tabular}[c]{@{}c@{}}Model D\\ $f=10^{6}$ GeV\end{tabular}} & $\rm D_1$ & 0.018      & 0.015      & 0.018         & 70698.9         & 23487.0         & 1.658    & 39613.7     \\ \cline{2-9} 
                                                                                      & $\rm D_2$ & 0.071      & 0.055      & 0.173         & 136258.8        & 50724.8         & 1.698    & 3261.5      \\ \cline{2-9} 
                                                                                      & $\rm D_3$ & 0.262      & 0.187      & 0.714         & 270259.0        & 176070.9        & 0.380    & 791.0       \\ \hline
    \end{tabular}
    \label{tab:i}
\end{table}

In Fig.~\ref{fig:GWsp} we plot the GW spectrum as a function of frequency, with various choices of parameters shown in Table~\ref{tab:i}.
The solid, dashed, and dash-dotted curves depict the spectra expected for model $\rm M_1$, $\rm M_2$, and $\rm M_3$, where M=A, B, C, and D.
The upper left, upper right, lower left and lower right plots assume the symmetry breaking scales of $f=10^3$~GeV, $10^4$~GeV, $10^5$~GeV, and $10^6$~GeV, respectively.
As shown in the plots, the peak frequency increases with the symmetry breaking scale. 
We observe that the $U(1)$ symmetry breaking scales in the range of $10^3-10^6$~GeV are all within the probe of the BBO and ALIA interferometers.
The LISA and Taiji interferometers can probe lower frequencies of GW signals, corresponding to the symmetry breaking scale $f\lesssim 10^4$~GeV.
On the other hand, the GW signals from scale of $f=10^6$~GeV could be detected by the ground-based GW observatories ET and CE.

\subsection{LIGO searches on stochastic gravitational waves}

\begin{figure}
    \centering
    \includegraphics[width=110mm,angle=0]{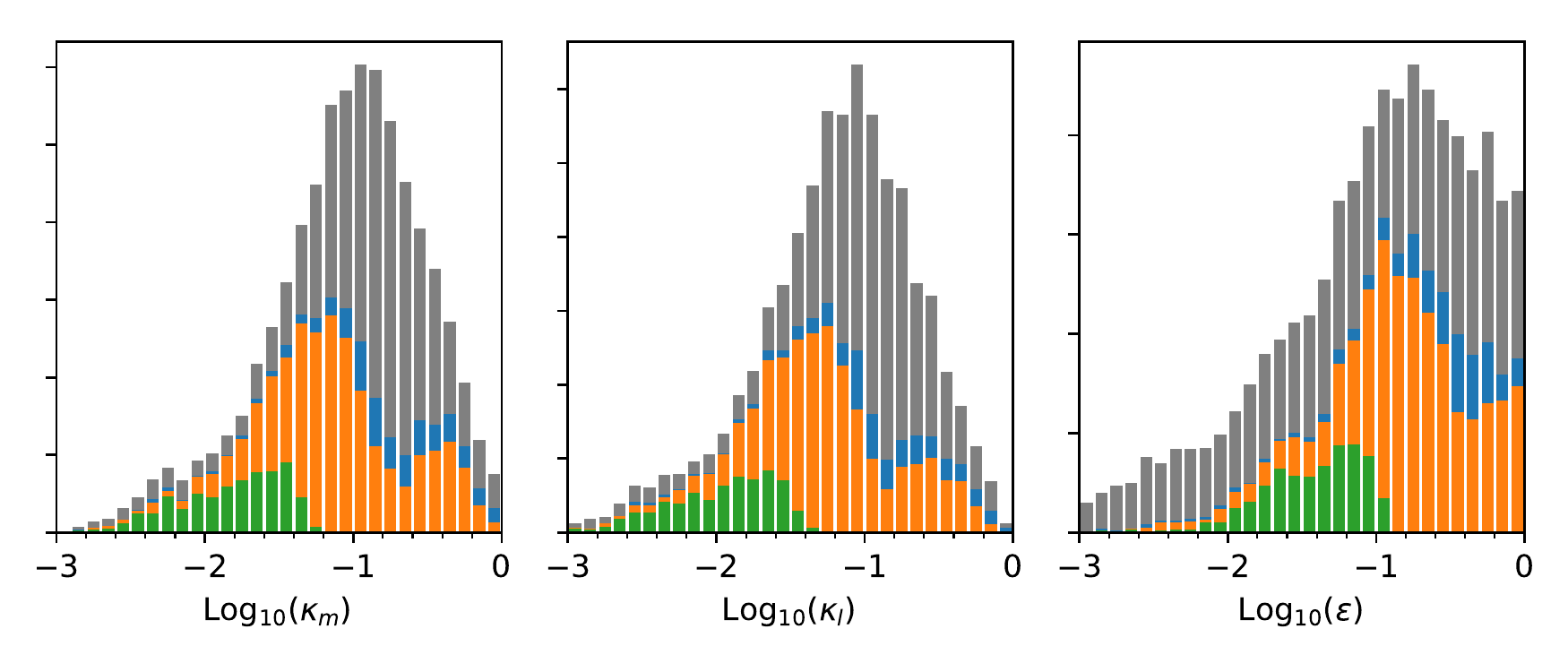}\\
    \includegraphics[width=110mm,angle=0]{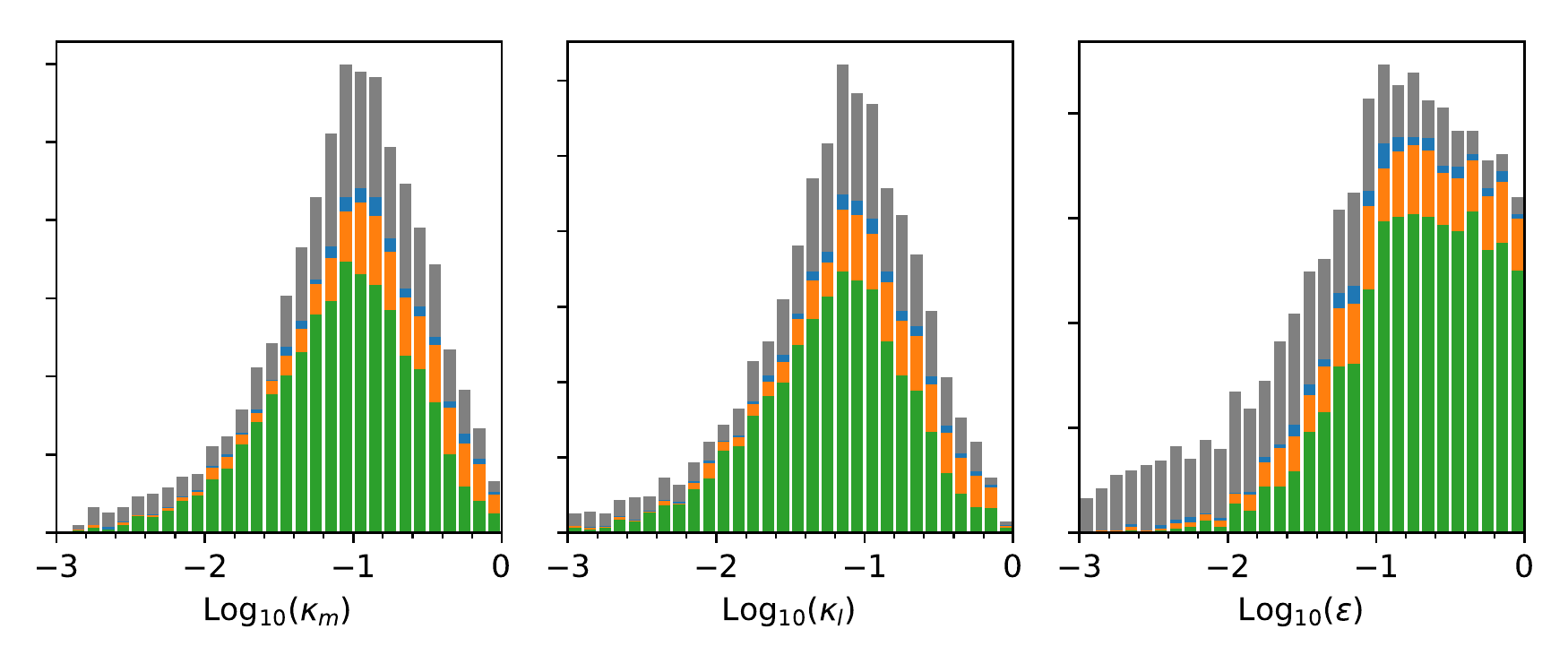}
    \caption{Histograms of the parameters that satisfy the nucleation condition.  The grey histograms represent the total samples, and the blue, orange, and green histograms denote those samples that can be probed for LIGO design, O3, and O2 run, respectively.  The top (bottom) row assumes the symmetry breaking scale to be $10^5$~GeV (and $10^6$~GeV).}
    \label{fig:dist_LIGO}
\end{figure}

We have shown that for the symmetry breaking scale at $f\sim 10^{6}$~GeV, the peak frequencies of the GW signals are right in the range that is covered by the ground-based GW observatories.
Searches for the isotropic stochastic GWs background have been undertaken by the LIGO and Virgo Collaborations. The results from a cross-correlation analysis of data from the two observing run phases (O1 and O2) of Advanced LIGO are shown in refs.~\cite{LIGOSGW2018PRL,LIGOSGW2019PRD}. 
The upper limits on the normalized energy density in GWs at the 95\% confidence level of $\Omega_{\rm GW}<6.0\times 10^{-8}$ at 25 Hz for a flat background have been obtained due to no evidence for the existence of a stochastic background.

Here we take advantage of the results from the Advanced LIGO O2 running to put constraints on
the clockwork axion model.  We will also estimate the detection prospects for the on-going Advanced LIGO observing run, including the third phase (O3) and design phase.
The second observing run, O2, has ben completed in 2017 and ran for approximately 9 months.
Following ref.~\cite{LIGOSGW2018PRL}, we use a duration of 12 months for O3 run and assume 24 months for the design phase (2022+).  
The GW background may be detectable with a $\rm SNR=3$~\cite{LIGOSGW2018PRL}.

\begin{figure}
    \centering
    \includegraphics[width=75mm,angle=0]{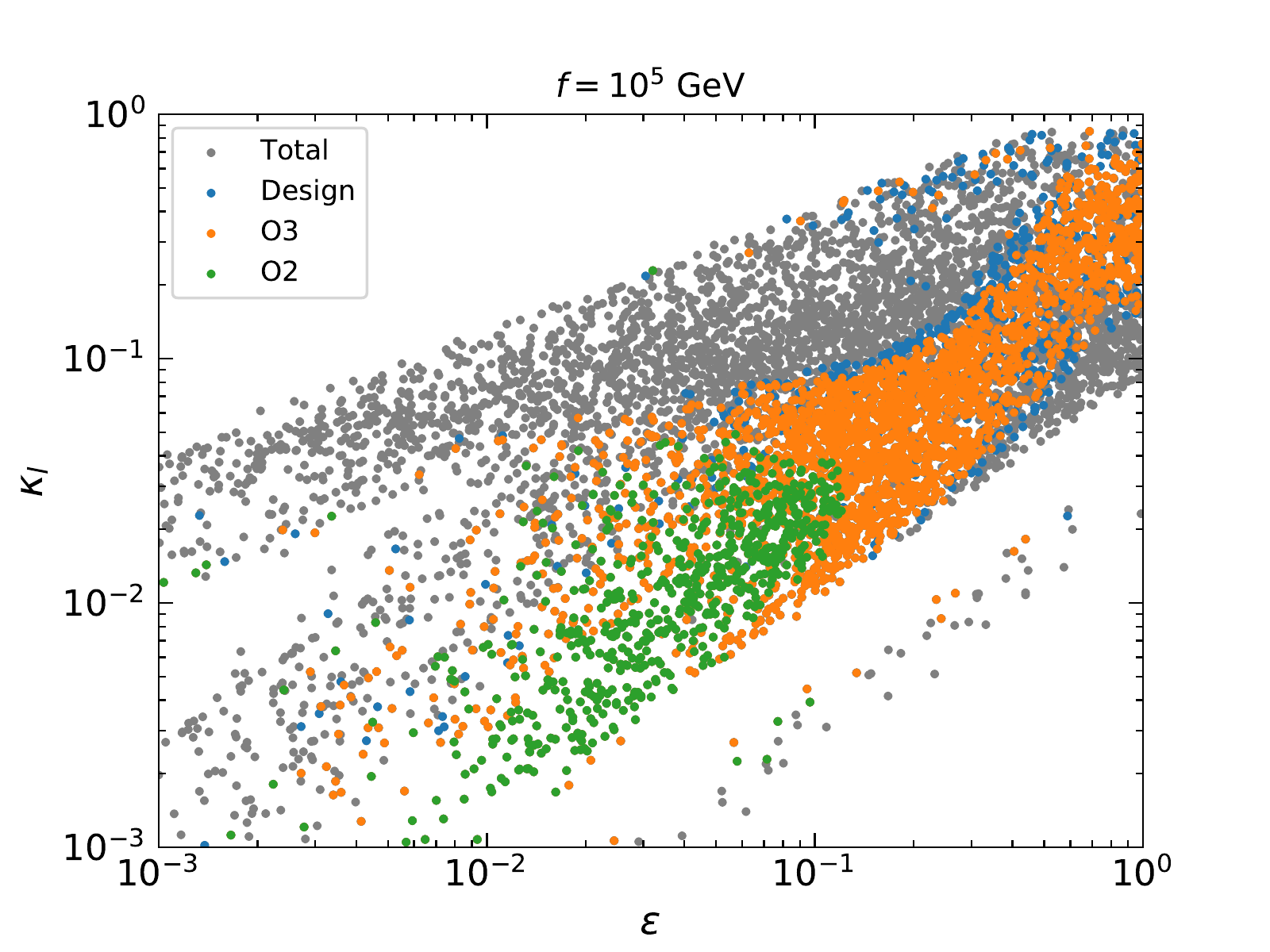}
    \includegraphics[width=75mm,angle=0]{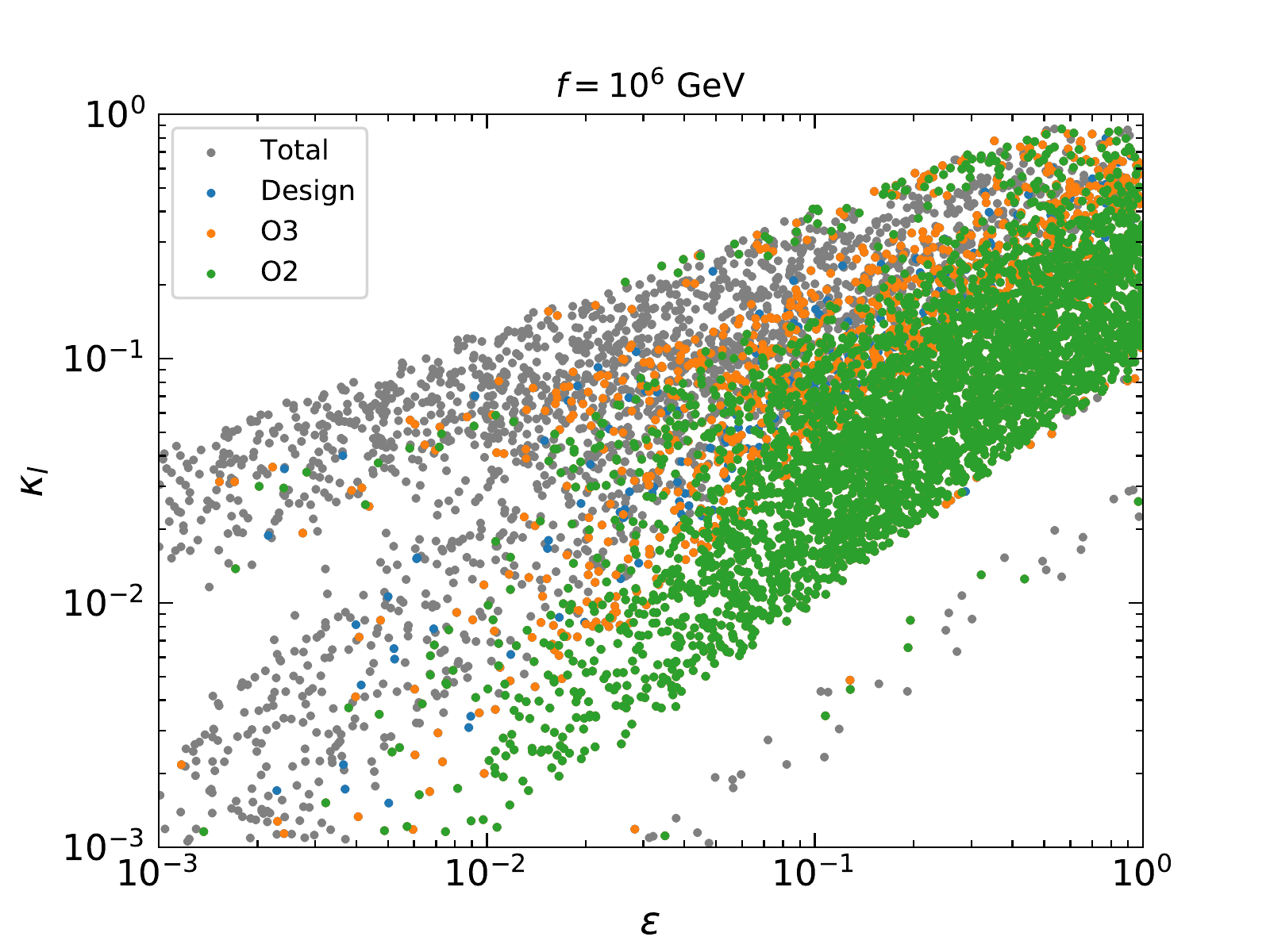}
    \caption{Scatter plots of the parameter distributions on the $\varepsilon-\kappa_l$ plane.
    The grey scatter dots represent the total samples that satisfy the nucleation condition, while the blue, orange, and green scatter dots denote those samples that can be probed by the LIGO design, O3, and O2 run, respectively.  We fix $f=10^5$~GeV and $10^6$~GeV in the left and right plots, respectively.}
    \label{fig:epkl}
\end{figure}

We calculate the GW signals from the vacuum phase transitions at all sites, assuming $N=10$, and compare them with the LIGO sensitivities.
The main results are shown in Fig.~\ref{fig:dist_LIGO} and Fig.~\ref{fig:epkl}.  The green, orange and blue histograms in Fig.~\ref{fig:dist_LIGO} represent detection ranges in the O2, O3 and design phases, respectively. The grey histograms are the total samples that can generate a successful phase transition.
We fix the symmetry breaking scale to $f=10^{5}$~GeV and $10^6$~GeV in the upper and lower plots of Fig.~\ref{fig:dist_LIGO}, respectively.
We further show the detectable regions for the LIGO run phases on the $\varepsilon-\kappa_l$ plane in Fig.~\ref{fig:epkl}.
Since no evidence for a stochastic GW background is found in the O2 run of LIGO, we interpret the green histograms in Fig.~\ref{fig:dist_LIGO} (and scatter points in Fig.~\ref{fig:epkl}) as those parameter regions
that have been constrained by the LIGO O2 observations.
Combine with Fig.~\ref{fig:dist_LIGO} and Fig.~\ref{fig:epkl}, we observe that for the symmetry breaking scale of $f=10^{5}$~GeV,
the parameter space with $\kappa_m\sim 0.06-0.001$, $\kappa_l\sim 0.04-0.001$, and $\varepsilon\sim 0.1-0.01$ has been excluded by LIGO O2 run.
Nearly half of the parameter space for $f=10^5$~GeV can be further tested by LIGO O3 and design phases.

\begin{figure}
    \centering
    \includegraphics[width=75mm,angle=0]{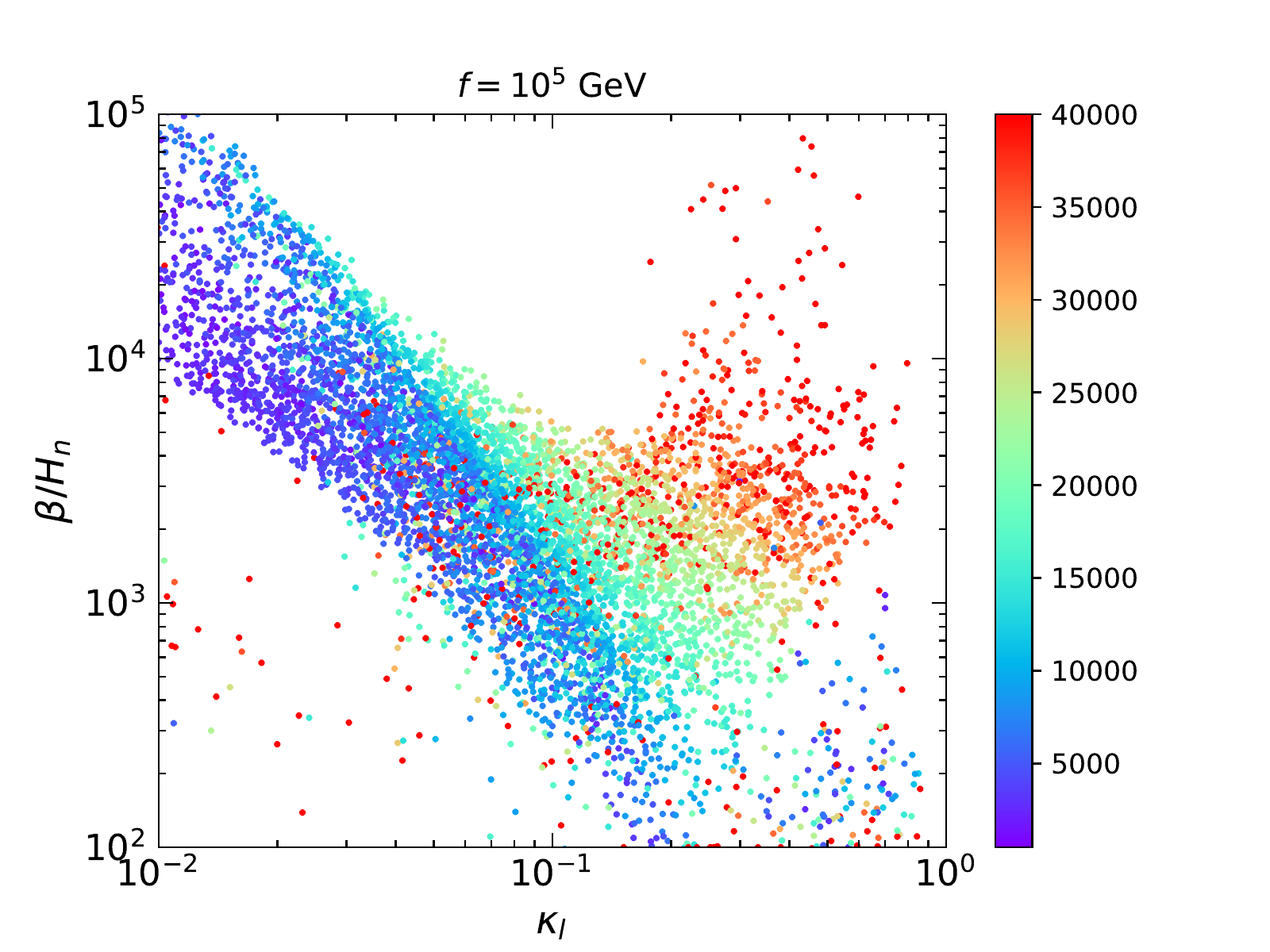}
    \includegraphics[width=75mm,angle=0]{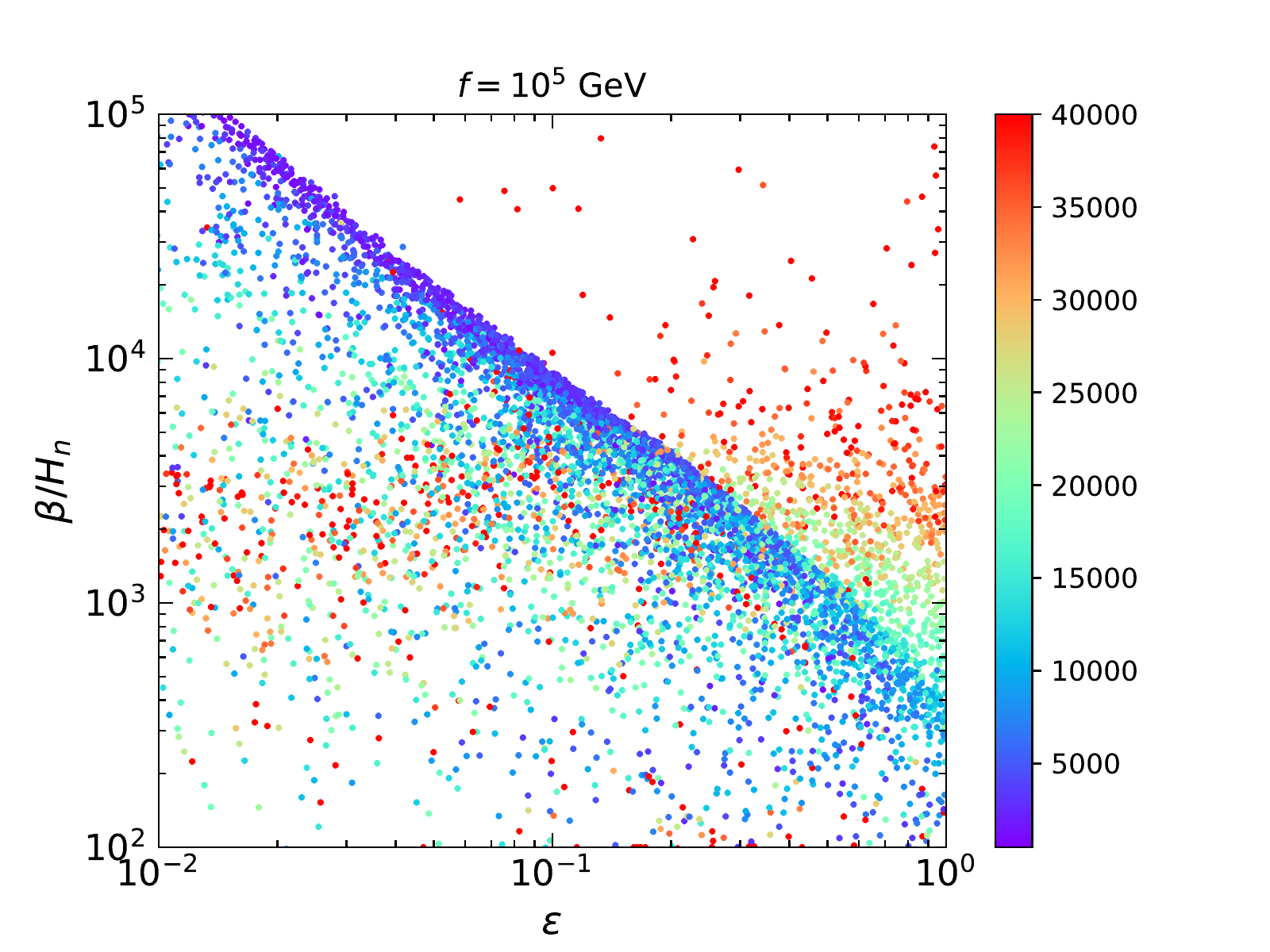}
    \caption{The distribution of $\beta/H_n$ as a function of $\kappa_l$ (left) and $\varepsilon$ (right).
    The colored bars represent the nucleation temperature. We fix $f=10^5$~GeV, and $10^6$~GeV in the left and right plot, respectively.}
    \label{fig:klepb}
\end{figure}

To see why lower values of $\kappa_m$, $\kappa_l$, and $\varepsilon$ for $f=10^5$~GeV can be effectively probed by the LIGO O2 run, we plot in Fig.~\ref{fig:klepb} the parameter $\beta/H_n$ as a function of the parameters $\kappa_l$ (left) and $\varepsilon$ (right), with the color indicating the nucleation temperature.  We observe that $T_n$ tends to increase with both $\kappa_l$ and $\varepsilon$, while $\beta/H_n$ tends to decrease as $\kappa_l$ and $\varepsilon$ increase.  For the parameters in the available range of O2 run, we find $T_n\lesssim 10^4$~GeV and $\beta/H_n\gtrsim 5\times 10^3$. 
With eq.~\eqref{eq:fturb}, the peak frequencies for these parameters fall in the range of $10-100$~Hz, which are the most sensitive frequency band 
for LIGO. However the amplitude of GW signal is suppressed by a large value of $\beta/H_n$ since $h^2\Omega_{\rm GW}$ is inversely proportional to
$\beta/H_n$ (with bubble collisions being neglected). For the case of $f=10^{6}$~GeV, we can achieve the LIGO sensitive frequency band with 
a higher nucleation temperature $T_n\gtrsim 10^5$~GeV and a lower $\beta/H_n\lesssim 5\times 10^{2}$.
Thus, the GW signals from symmetry breaking at the scale $f=10^{6}$~GeV can be sufficiently loud for LIGO O2 run, as indicated the Fig.~\ref{fig:dist_LIGO} and Fig.~\ref{fig:epkl}. We observe that for $f=10^{6}$~GeV, most of the parameter space has been excluded by 
while the LIGO O2 run, the remaining regions can be further tested by LIGO O3 and design phases.

\section{Gravitational waves from domain wall annihilation}
\label{sec:GWDW}

\subsection{Gravitational wave spectrum}

The network of cosmic strings and domain walls form after the phase transition of the clockwork axion model.
Numerical simulations~\cite{Higaki2016JHEPb} show that for large number of $N$ ($\gtrsim 3$), this string-wall network can survive until the QCD phase transition.  
The QCD instanton effects at $T\lesssim 1$~GeV give rise to the QCD axion potential, 
$V_{\rm bias}\sim \Lambda_{\rm QCD}^4$ (where $\Lambda_{\rm QCD}=(332\pm 17)$~MeV~\cite{Tanabashi2018PRD} is the the QCD confinement scale), 
which serves as an energy bias to break the degeneracy of discrete vacua, and thus, leads to the annihilation of the domian walls.

In the scaling regime for long-lived domain walls, the evolution of the energy density of domain walls can be parameterized as 
\begin{equation}
    \label{eq:rhowall}
    \rho_{\rm wall}(t)=\mathcal{A}\frac{\sigma}{t}
    ~,
\end{equation}
where $\sigma\simeq 8m_A^2f^2$ is the tension of the domain wall~\cite{Higaki2016JHEPb,Long2018JHEP}, $t=1/(2H)$, and $\mathcal{A}\sim 1.0$ 
from the analysis of numerical simulations~\cite{Kawasaki2015PRD}. The annihilation of domain walls becomes significant when the tension of domain walls is 
comparable with the volume pressure $p_{\rm V}\sim V_{\rm bias}$, and the annihilation temperature of domain walls is given by~\cite{Saikawa2017}
\begin{equation}
    T_{\rm ann}\simeq 7.15\times 10^{-2}~{\rm GeV}~\varepsilon^{-1/4}\left(\frac{g_{*}\left(T_{\mathrm{ann}}\right)}{10}\right)^{-1/4}
    \left(\frac{f}{100~\rm{TeV}}\right)^{-3/2}\left(\frac{\Lambda_{\rm QCD}}{100~\rm{MeV}}\right)^{2}.
\end{equation}

Eq.~\eqref{eq:rhowall} shows that the energy density of domain walls in the scaling regime decreases as $\propto t^{-1}$.  This decay rate is slower than those of dusts $\propto t^{-3/2}$ and radiation $\propto t^{-2}$.
From the condition $\rho_c=\rho_{\rm wall}$, where $\rho_c$ is the critical density of the Universe, the energy density of the Universe 
would eventually be dominated by the domain walls at the temperature
\begin{equation}
    T_{\rm dom}=5.44\times 10^{-2}~{\rm GeV}~\varepsilon^{1/4}\left(\frac{g_{*}\left(T_{\mathrm{ann}}\right)}{10}\right)^{-1/4}
    \left(\frac{f}{100~\rm{TeV}}\right)^{3/2}.
\end{equation}
Requiring the domain walls annihilate before they dominate the Universe, {\rm i.e.,} $T_{\rm ann}\gtrsim T_{\rm dom}$, we have
\begin{equation}
    f\lesssim 100~{\rm TeV}~\varepsilon^{-1/6}\left(\frac{\Lambda_{\rm QCD}}{100~\rm{MeV}}\right)^{2/3}.
\end{equation}
We thus find an upper bound on the symmetry breaking scale, $f\lesssim 400$~TeV.

The production of GWs from the annihilation of domain walls has been studied in the literature.
The peak amplitude of the GW spectrum is produced at the annihilation time of domain walls, 
\begin{equation}
    \Omega_{\mathrm{GW}}\left(\nu_{\mathrm{peak}}\left(t_{\mathrm{ann}}\right)\right)
    \simeq \frac{8 \pi \tilde{\epsilon}_{\mathrm{gw}} G^{2} A^{2} \sigma^{2}}{3 H_{\mathrm{ann}}^{2}}
    ~,
\end{equation}
where the efficiency of the gravitational wave emission $\tilde{\epsilon}_{\rm{gw}} \simeq 0.7 \pm 0.4$~\cite{Hiramatsu2014JCAP}
and $A\simeq N$ from the numerical simulations~\cite{Higaki2016JHEPb}.
With the expansion of the Universe, the amplitude is diluted as $\propto R(t)^{-4}$, where $R(t)$ is the scale of the Universe.
The peak amplitude of the GW spectrum today is given by
\begin{equation}
    h^2\Omega_{\rm{GW}}^{\rm peak}\left(t_{0}\right)
    =6.45 \times 10^{-6} \varepsilon \left(\frac{\tilde{\epsilon}_{\rm{gw}}}{0.7}\right )\left( \frac{A}{10} \right )^2
    \left(\frac{g_{* s}\left(T_{\rm{ann}}\right)}{10}\right)^{-4/3}\left(\frac{f}{100~\rm{TeV}}\right)^{6}
    \left(\frac{T_{\rm{ann}}}{0.1~\rm{GeV}}\right)^{-4},
\end{equation}
with the red-shifted peak frequency from $\nu_{\rm{peak }}\left(t_{\rm{ann }}\right) \simeq H_{\rm{ann }}$ given by
\begin{equation}
    \label{eq:vpeak}
    \nu_{\rm peak}(t_0)\simeq 1.1 \times 10^{-8} \mathrm{~Hz}\left(\frac{g_{*}\left(T_{\mathrm{ann}}\right)}{10}\right)^{1/2}
    \left(\frac{g_{*s}\left(T_{\mathrm{ann}}\right)}{10}\right)^{-1 / 3}\left(\frac{T_{\mathrm{ann}}}{0.1~\rm{GeV}}\right).
\end{equation}
The analysis of numerical simulations in ref.~\cite{Hiramatsu2014JCAP} shows that the frequency dependence of the GW spectrum can be approximately parameterized as 
\begin{equation}
    h^2\Omega_{\rm GW}=
    \begin{cases}
    \displaystyle
    h^2\Omega_{\rm GW}^{\rm peak}\left ( \nu \over \nu_{\rm peak} \right )^3~{\rm for~\nu<\nu_{\rm peak}}
    \\ 
    \displaystyle
    h^2\Omega_{\rm GW}^{\rm peak}\left ( \nu_{\rm peak} \over \nu \right )~~{\rm for~\nu>\nu_{\rm peak}}.
    \end{cases}
\end{equation}

\subsection{NANOGrav pulsar timing observations}

The byproducts of the clockwork axion phase transition, the domain walls, dominantly annihilate at a temperature around $T_{\rm ann}\sim 0.1-1$~GeV, right after the QCD confinement.  From eq.~\eqref{eq:vpeak}, we see that the peak frequency of GWs from domain wall annihilation is around $\sim 10^{-8}$~Hz, falling in the frequency band probed by pulsar timing observations.
The searches for nHz isotropic stochastic GW background via the observation of pulsars are performing by, for example, the European Pulsar Timing Array (EPTA)~\cite{EPTA2015} and the NANOGrav~\cite{NANOGrav2018}, 
over a long time span. The constraints from the observations of 18 years by EPTA and over 11 years by NANOGrav have restricted the GW background amplitude in the frequency band of $2-10$~nHz to be $h^2\Omega_{\rm GW}\lesssim 5\times 10^{-9}$ and $10^{-9}$, respectively.
These results constrain the symmetry breaking scale to be $f\lesssim 300$~TeV, slightly tighter than the constraints from requiring domain walls to annihilate before they dominate the energy density of the Universe.

\begin{figure}
    \centering
    \includegraphics[width=75mm,angle=0]{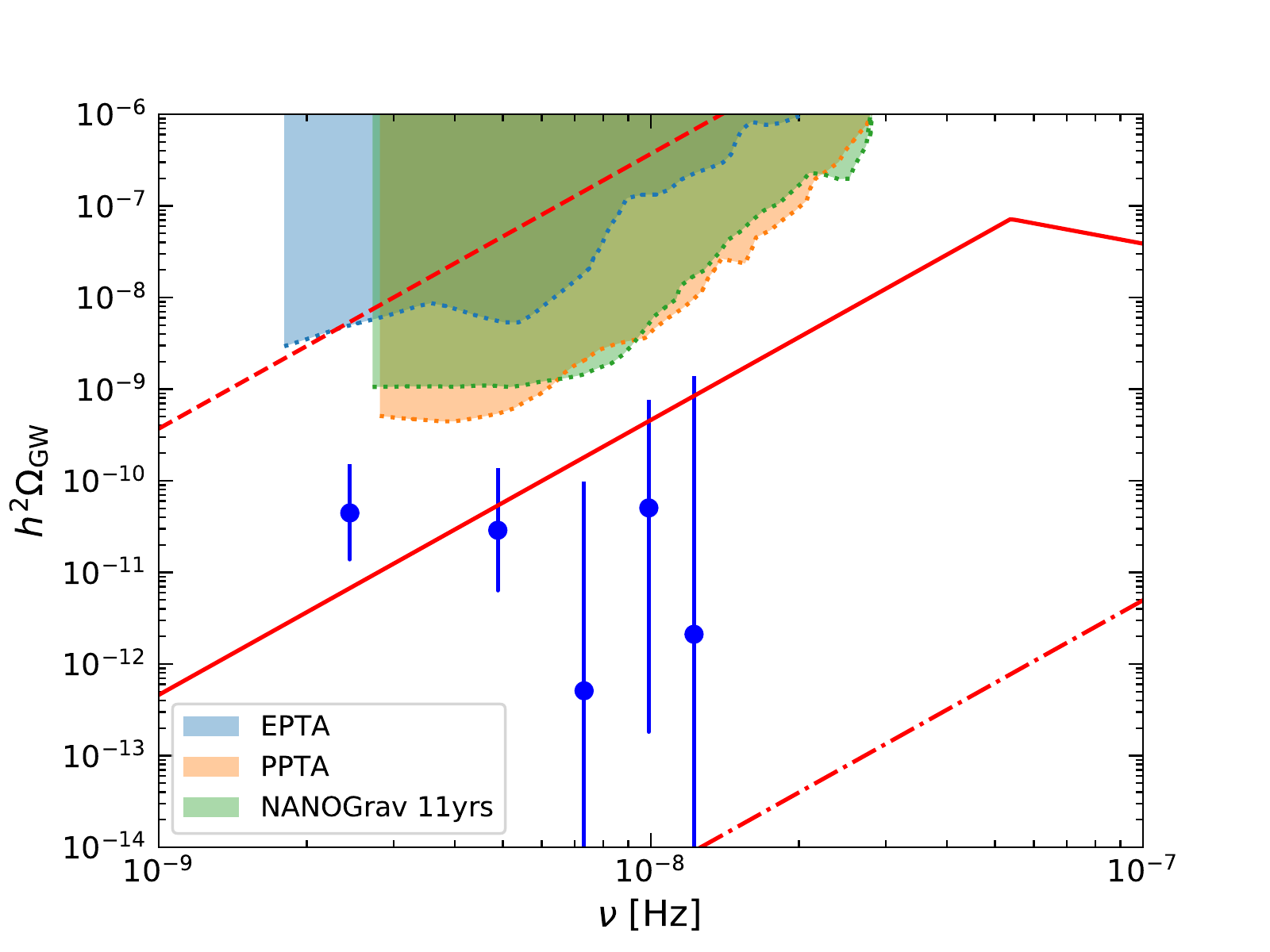}
    \includegraphics[width=75mm,angle=0]{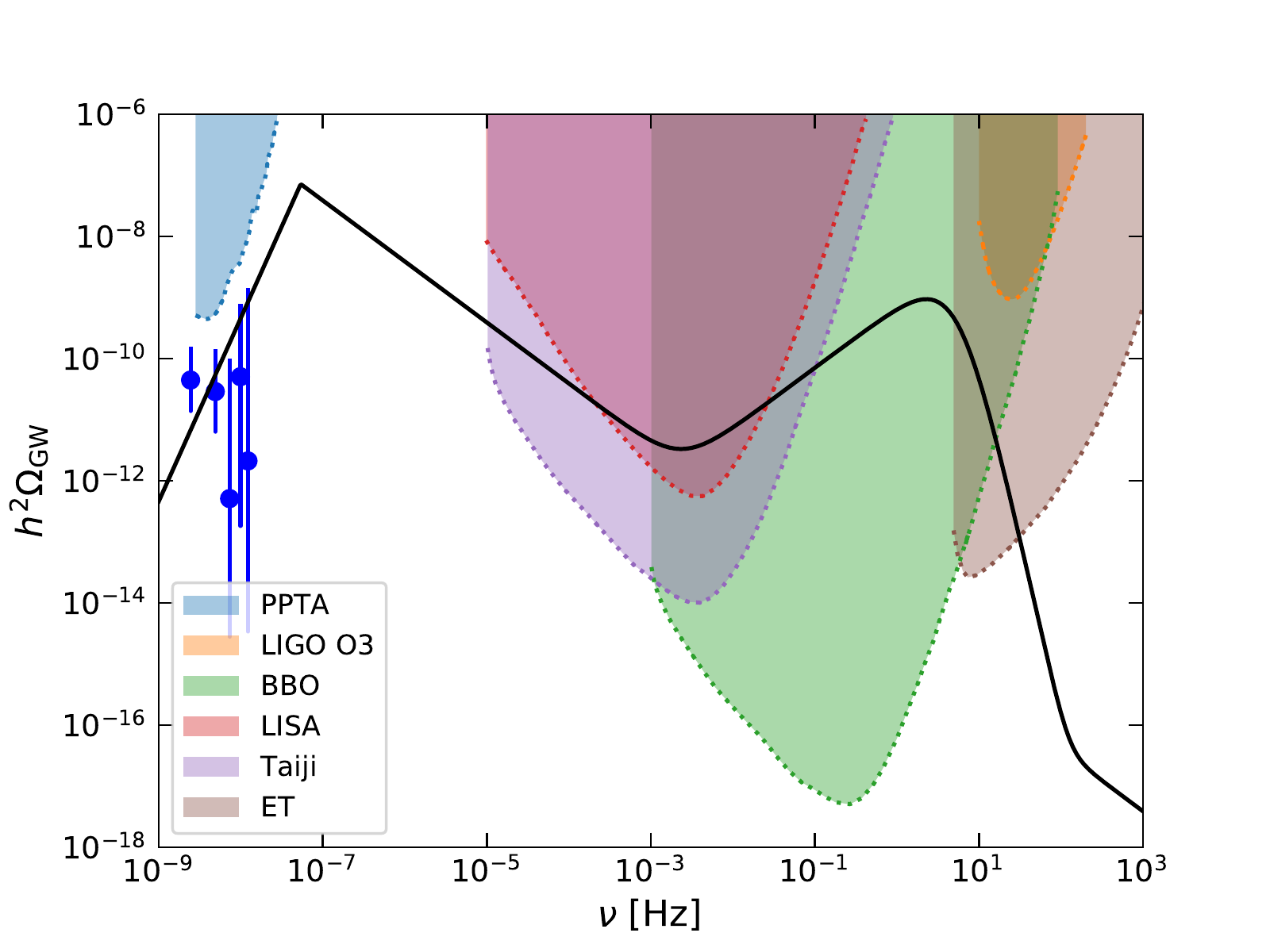}
    \caption{The GW spectra and various experimental sensitivities as a function of frequency. In both plots, the blue data points represent
    the analysis of NANOGrav 12.5-year data, the parameter is fixed at $\kappa_m=0.030$, $\kappa_l=0.024$, and $\varepsilon=0.1$, respectively.
    In the left plot, the red dashed, solid, and dashed-dot lines represent the GW spectrum from domain wall annihilation with a symmetry 
    breaking scale $f=300$~TeV, $f=200$~TeV, and $f=100$~TeV, respectively. 
    The black line in the right plot shows the total GW signal from domain wall annihilation and first-order phase transition.}
    \label{fig:NANOGrav}
\end{figure}

The NANOGrav Collaboration has recently released their analysis on the 12.5-year data~\cite{NANOGrav2020}
and one signal of a stochastic spectrum was found within the frequency band $\sim 1-10$~nHz.
Various possible sources of the signal have been proposed, including cosmic strings~\cite{Ellis2020NANO,Blasi2020,Buchmuller2020PLB,Samanta2020}, first-order cosmological phase transitions~\cite{Nakai2020,Addazi2020,Neronov2020,Bian2020,Li2020,Paul2020}, coherent oscillation of axionic fields~\cite{Ratzinger2020,Namba2020}, and primordial black holes~\cite{Vaskonen2020,Vaskonen2020,Domenech2020}.
As indicated by eq.~\eqref{eq:vpeak}, the nHz stochastic GWs can naturally be produced by the annihilation of domain walls in the clockwork axion model~\cite{Higaki2016JHEPb}.
In the left plot of Fig.~\ref{fig:NANOGrav}, we show the GW relic energy density from the domain wall annihilation, taking $\varepsilon=0.1$.  The red dashed, solid, and dash-dotted lines represent the results with a symmetry breaking scale of $f=300$~TeV, $f=200$~TeV, and $f=100$~TeV, respectively.  We find that GWs from the model with a scale $f=200$~TeV can explain the NANOGrav 12.5-year data in blue points.
In addition to the nHz GW signals, the first-order phase transition of the clockwork work axion model at the scale of $f=200$~TeV can also produce a GW signal peaked around $\sim 10$~Hz.  We show the result in this scenario with the black line in the right plot of Fig.~\ref{fig:NANOGrav}.
As we have shown in the previous section, most of the parameter space for the first-order phase transition at the scale $f=10^{3}$~TeV has been excluded by the LIGO O2 run. The O3 and the design phases of LIGO could probe half of the parameter space of the phase transition at the scale around $10^{2}$~TeV.  We thus expected another signal in the LIGO O3 run if the NANOGrav 12.5-year data are indeed induced by the annihilation of
domain walls from the phase transition of the clockwork work axion at the scale of $f=200$~TeV.

\section{Conclusions}
\label{sec:summary}

We have shown in this work the opportunity to explore the clockwork axion with the information from gravitational wave (GW) detection.
It is well known that GWs can be produced from violent first-order cosmological phase transitions.  
However, in the conventional QCD axion models the PQ phase transition is second-order. We show that the PQ phase transition can be first-order when the dimension-6 operator 
is included in the scalar potential. Based on a comprehensive scan in the parameter space, we find that the parameters $\kappa_m$, $\kappa_l$, and $\varepsilon$ have been restricted within the range of $10^{-2}-1.0$ in order to trigger a first-order phase transition while having successful nucleation of the true vacuum bubble.
We show that the GWs from the PQ phase transition at scales in the range of $10^3-10^6$~GeV can be probed by the BBO and ALIA interferometers.
The LISA and Taiji interferometers could probe GW signals of lower frequencies, corresponding to the symmetry breaking scale of $f\lesssim 10^4$~GeV.
On the other hand, the GW signals from the scale $f\gtrsim 10^5$~GeV could be detected by the ground-based GW observatories ET and CE.
In fact, we find that for the symmetry breaking scale of $f=10^{5}$~GeV, the parameter space of $\kappa_m\sim 0.06-0.001$, $\kappa_l\sim 0.04-0.001$, and $\varepsilon\sim 0.1-0.01$ has been excluded by LIGO O2 run. Nearly half of the parameter space for $f=10^5$~GeV can be further tested by the LIGO O3 and design phases.  For the scale of $f=10^{6}$~GeV, we find that most of the parameter space has been excluded by the LIGO O2 run, with
the remaining regions further testable by the LIGO O3 and design phases.

The QCD axion potential arises after the QCD confinement can serve as the bias potential to annihilate the domain walls produced from the PQ phase transition.  We show that the PQ scale $f$ should $\lesssim 4\times 10^{5}$~GeV to ensure that the domain walls annihilate before they dominate the energy density of the Universe.
We find that the GWs from the annihilation of domain walls with the scale $f=2\times 10^5$~GeV can induce the stochastic GW background signals indicated by the NANOGrav 12.5-year observation. The running of the LIGO O3 phase has a chance to observe the GW signals from the first-order PQ phase transition at this scale in the clockwork axion model.  The future space-based GW interferometers, including LISA, Taiji, ALIA, DECIGO, and BBO, the ground-based GW experiments, ET and CE, as well as the pulsar timing arrays, such as PPTA~\cite{Hobbs2013}, IPTA~\cite{Hobbs2013}, and SKA~\cite{Janssen2015} will be able to further provide more 
opportunities to test the clockwork axion model.

\acknowledgments{
BQL thanks Da Huang for helpful comments and suggestions.  This work was supported in part by the Ministry of Science and Technology (MOST) of Taiwan under Grant Nos.~MOST-108-2112-M-002-005-MY3 and 109-2811-M-002-550. 
}

\bigskip

\end{document}